\documentclass[
12pt,
]{article}

\usepackage{amssymb}
\usepackage{amsthm}
\usepackage{amsmath}

\theoremstyle{plain}
  \newtheorem{thm}{Theorem}[section]
  \newtheorem{lem}{Lemma}[section]
  \newtheorem{cor}{Corollary}[section]
  
\theoremstyle{definition}
  \newtheorem{defn}{Definition}[section]

\theoremstyle{remark}
  \newtheorem{rem}{Remark}

\newcommand{\Cal}[1]{\ensuremath{\mathcal{#1}}}
\newcommand{\g}[1]{\mbox{\boldmath ${#1}$}}

\newcommand{\R}{\Bbb{R}}

\newcommand{\p}{\partial}

\newcommand{\ve}{\varepsilon}
\newcommand{\dai}{\mathop{\mathrm{supp}}}

\numberwithin{equation}{section}

\date{\today}

\title{
Time-dependent methods in 
inverse scattering problems 
 for the Hartree-Fock  equation}

\author{Michiyuki Watanabe 
\thanks{Department of Education, Niigata University, 
8050, Ikarashi2-no-cho, Nishi-ku, Niigata City 950-2181, Japan
(email: {\tt mwatanab@ed.niigata-u.ac.jp}).}
}

\begin{document}

\maketitle

\begin{abstract}
The inverse scattering theory for many-body systems in quantum mechanics is an 
important and difficult issue not only in physics---atomic physics, molecular physics and 
nuclear physics---but also mathematics. 
The major purpose in this paper is to establish a reconstruction procedure of 
two-body interactions from scattering solutions for a Hartree-Fock equation. 
More precisely,  
this paper gives a uniqueness theorem and 
proposes a new reconstruction procedure of the short-range and two-body interactions 
from a high-velocity limit of the scattering operator for the Hartree-Fock equation. 
Moreover, it will be found that the high-velocity limit of the scattering operator is equal to 
a small-amplitude limit of it. 
The main ingredients of mathematical analysis in this paper are based on 
the theory of integral equations of the first kind  
and a Strichartz type  estimates on a solution to the free Schr\"{o}dinger equation. 
\end{abstract}

\begin{quote}
{\small 
 \begin{description}
  \item[Keywords.] 
Hartree-Fock equations,  Non-linear Schr\"{o}dinger equations, 
$N$-body systems, Inverse scattering problems. 
  \item[2010 Mathematics Subject Classification.] 
Primary: 35P25; Secondary: 81U40, 35R30. \end{description}
 }
\end{quote}

\section{Introduction}
\subsection{Background}
Inverse scattering problems in quantum many-body systems are  important and difficult 
problems not only quantum physics but also mathematics. 
Some results on inverse 
scattering problems for $N$-body Schr\"{o}dinger equations 
were investigated. 
A reconstruction problem of identifying the two-body interactions from the 
high-energy asymptotics was studied in 
Wang \cite{wang},  
Enss and Weder \cite{Enss-Weder1995}, 
Novikov \cite{novikov} and 
Vasy \cite{vasy}.   
Uhlmann and Vasy \cite{uhlmann-vasy} studied a low-energy inverse scattering problem. 

As is well known, the solution of the $N$-body Schr\"{o}dinger equation on $\R^n$ is a 
high-dimensional complicated function on $\R^{nN}$, which usually causes exact or numerical 
calculations impractical. 
Therefore, methods of approximation in understanding the many-body problem in quantum 
mechanics have most often been proposed. 
A result on inverse scattering problems in nuclear physics by using the optical model, 
which is one of the method of approximation in the many-body problems, 
was reported by Isozaki-Nakazawa and Uhlmann \cite{isozaki-nakazawa-uhlmann}. 

The time-dependent Hartree-Fock approximation, which is one of the simplest approximate theories 
for solving the many-body Hamiltonian, has received much attention  due to its effect of 
calculations and having a wide field of applications 
(see, e.g., Goeke and Reinhard \cite{goeke-reinhard},  and 
Kramer and Saracero \cite{kramer-saraceno}). 
Time-dependent Hartree equations have also play an important role in the development of mathematical analysis 
due to its non-linear structure, 
which cause interesting behavior of the solutions (see, e.g., Cazenave \cite{Cazenave2003}). 

In this paper, we are interested in an inverse scattering problem for 
a time-dependent Hartree equation and a Hartree-Fock equation. 
Consider the $N$-body system of identical particles with the interaction potential $V_{\text{int}}$ 
consisting of sum of two-body force: 
$V_{\text{int}}=\sum_{i<j}V(x_i-x_j)$. 
Here we denoted the position of the $j$-th particle by $x_j$. 
The indistinguishability of identical particles permits that 
the interaction potential is symmetric: $V(x)=V(-x)$ for $x\in \R^n$.  

Put
 \begin{align*}
     V_H (x, u) u_j (x)& = \left\{ \int_{\R^n} 
     V(x-y) \sum_{\substack{k=1 \\ k \not=j}}^N 
      | u_{k}(y) |^2 \, dy \right\} 
     u_{j}(x), \\
     &=
     \left(V* \sum_{\substack{k=1 \\k\not=j}}^N | u_k |^2 \right) u_j(x),  \qquad u=\{u_j \}_{1\le j \le N }
 \end{align*}
and 
 \begin{align*}
   V_F(x,y)= - V(x-y) \sum_{\substack{k=1 \\k\not=j}}^N \overline{u_k} (y) u_k (x). 
 \end{align*}
then the Hartree equation (H equation) is written as  
 \begin{equation}\label{eqn:1-1}
  i \frac{\p u_j }{\p t} = H_0 u_j + V_H ( x, u ) u_j \qquad \text{for $1\le j \le N$}
 \end{equation}
and the Hartree-Fock equation (HF equation) is written as 
 \begin{equation}\label{eqn:1-2}
  i \frac{\p u_j }{\p t} = \{H_0  + V_H ( x, u ) \} u_j  
   + \int_{\R^n} V_F (x,y) u_j (y) \, dy
   \qquad \text{for $1\le j \le N$},
 \end{equation}
where $H_0= -\dfrac{1}{2} \Delta$ and $u_j= u_j(t,x)$ is an unknown function in 
$(t,x) \in \R \times \R^n$. 
The terms  $V_H( x,u ) u_j (x)$  and  $\int V_F(x,y) u_j (y) dy$ 
are called the Hartree term and the Fock term, respectively.  

The problem considered in this paper is to reconstruct the interaction potential $V(x)$ from the 
corresponding scattering operator defined below.

The H equation and the HF equation are non-linear Schr\"{o}dinger equations with cubic 
convolution non-linearity. 
Thus, our inverse problems can be labeled as the inverse non-linear scattering problems of identifying the 
non-linearity from the scattering operator. 
As is well-known, inverse scattering problems are non-linear problems even if 
governing differential equations are linear equations. 
From an analytical point of view, inverse problems of non-linear differential equations are 
one of the most difficult problems in inverse problems.

Initial attempts for inverse non-linear scattering problems focused on 
identifying the coefficients of power type non-linearity from 
the small scattering data. 
The approach called small-amplitude method was developed by Strauss \cite{Strauss1974},  
Weder \cite{Weder1997, Weder2000_1, Weder2000_2, Weder2001_1, Weder2001_2, Weder2002, Weder2005_1} and 
Angeles-Romero-Weder \cite{Angeles-Romero-Weder2006}, 
which has been shown to be powerful to reconstruct coefficient functions 
of  non-linear Schr\"{o}dinger equations. However, the approach is valid only for small data. 
Reconstruction of  coefficient functions from large scattering data requires  
an alternative approach. 
Recently, in \cite{watanabe18}, the author establishes the unique reconstruction of the power type 
non-linearity from the large scattering data by using the method of high-velocity limit  
developed by Enss and Weder \cite{Enss-Weder1995}.

On the inverse scattering problem for Hartree equations, 
most references we know are concerned with 
uniqueness results \cite{sasaki-watanabe} and 
reconstructions for interactions with special form 
\cite{watanabe0, watanabe2, watanabe2002, watanabe2007-1}---all focused on identifying the 
interactions from the small scattering data. 

With regard to the inverse scattering problem for Hartree-Fock equations, 
little work has been done even for a uniqueness problem.  
The only reference we know is the work \cite{watanabe2007} 
where the reconstruction formula is given for the special interactions of the form  
$V_j(x)=\lambda_j |x|^{-\sigma_j}$, $1\le j \le 3$ in the case of 3-body systems. 

In this paper, we deal with the inverse scattering problem for both  
H equation \eqref{eqn:1-1} and HF equation \eqref{eqn:1-2}.    
This paper presents a explicit reconstruction formula on recovering the two-body 
interaction  $V(x)$ 
from the high energy asymptotics of the scattering solutions for 
the H equation \eqref{eqn:1-1} and the HF equation \eqref{eqn:1-2}, respectively. 
A uniqueness theorem in the inverse scattering problem for the HF equation is 
also proved. 
As is mentioned above, the fundamental ingredients in mathematical analysis to investigate the non-linear 
inverse scattering problems are now two methods---the small-amplitude method  
and the high-velocity method.  
This paper will uncover the relation between the two methods.

\subsection{Methods} 
Our method consists of the following analysis:  
\begin{itemize}
 \item Asymptotic expansion of the scattering operator acting on the function 
 $\Phi_v(x) = e^{iv\cdot x}\varphi (x)$ as $|v|\to \infty$. 
 
 \item Derivation of  a transformation of the Fourier transform of the interaction potential 
 $\widehat{V}(\xi)$ of the form $\int G(\xi, \lambda) \widehat{V}(\xi)d\xi$ 
 by using a scale transform $\varphi_{\lambda}(x)= \varphi((\lambda+1)x)$. 
 
 \item Invertibility of the transformation by using the Picard's theorem 
 for an equation of the first kind with a compact operator. 
\end{itemize}

Previous researches \cite{Enss-Weder1995, watanabe18} explain that 
the high-velocity analysis of the scattering operator gives the Radon transform of the 
unknown coefficient functions. 
Due to the inversion formula of   
the Radon transform, 
the high-velocity analysis therefore provides  a reconstruction formula for  
unknown coefficient functions.  

With regard to Hartree equations, however, high-velocity analysis  
gives a transformation which is different from the Radon transform---the transformation   
has a complicated integral kernel due to those non-linearities. 
Invertibility of the transformation was unclear.

In order to overcome this difficulty, we employ the high-velocity analysis 
with scale transform, which leads an integral equation of the first kind for 
the unknown interaction $V$. 
It will be shown that 
the integral kernel is equicontinuous and equibounded in some function spaces. 
The Ascoli's lemma  therefore gives the compactness of the integral operator. 
Then, the Picard's theorem for an equation with a compact operator can give 
an explicit solution of the integral equation. 
We also remark that the Picard's theorem dose not imply uniqueness of the 
solution. 
Construction of the proper initial data for the free Schr\"{o}dinger equation 
such that the non-linear interactions of the free solution can be localized at any 
fixed point in $\R^n$ lead us to prove a uniqueness theorem. 
This is done in Section 4.

It should be mentioned that 
the Picard's theorem can be applicable on a Hilbert space. 
This paper finds the proper Hilbert space to apply the Picard's theorem 
and consequently the proper function space to reconstruct the interaction $V$.

The fundamental ingredient in our proof  is a time-space $L^4$ estimate 
on a solution to the free Schr\"{o}dinger equations. 

\subsection{Results}
We summarize our main results. 
Let $W^{k,l}(\R^n)$ be the usual Sobolev space in $L^l(\R^n)$. 
We abbreviate $W^{k,2}(\R^n)$ as $H^k(\R^n)$. 

We first state results for the restricted Hartree equation (RH equation), 
which is the case for $N=2$ with $u_1=u_2$ in equation \eqref{eqn:1-1}.  
Proofs of theorems for H equation and HF equation are reduced to the proof of 
theorems on RH equation.

\subsubsection{Restricted Hartree equation}\label{subsub1-3-2}

Consider the RH equation:

\begin{equation}\label{restricted Hartree}
 i\p_t u = H_0 u +V_H ( x, u )u,
\end{equation}
where 
$V_H ( x,u )u=(V*|u|^2)u$. 
In order to formulate our inverse problem, let us state a result on the large date scattering 
problem. 
We denote solutions $u(t):=u(t,x)$ of  the equation \eqref{restricted Hartree} with initial data 
$f$  as $U(t)f$ and  
the unitary group of the self-adjoint operator 
$H_0=-\frac{1}{2}\Delta$ with a domain $H^1(\R^n)$ as 
$U_0(t)$.

\begin{thm}[Nakanishi \cite{nakanishi1999}]\label{thm:1-1}
Let $n\ge 3$. 
Assume that 
$V(x)$ is a radial, non-negative and non-increasing function such that 
\[ V\in  \Cal{V} := \left\{ V\in L^{p_1} + L^{p_2} \, ; \, 
 p_1,  \,  p_2 \ge 1, \,   n/2 > p_1 \ge  p_2 > n/4 
  \right\}.
\]
Then for any $f_- \in H^1(\R^n)$, there exists a unique pair of functions 
$f_+ \in H^1(\R^n)$ and $\varphi \in H^1(\R^n)$ such that 
\[ \| U(t)\varphi -U_0(t)f_{\pm}\|_{H^1}\longrightarrow 0, 
\qquad \text{as $t \to \pm \infty$}.
\]
In addition, the scattering operator 
\[ S: \, H^1(\R^n) \ni f_- \longrightarrow f_+\in H^1(\R^n)\]
is  homeomorphism in $H^1(\R^n)$.
\end{thm}

We term solutions constructed in Theorem \ref{thm:1-1} scattering solutions. 

\begin{rem}
The statement  in \cite[Theorem 1.1]{nakanishi1999} assumes that 
$p_1, p_2 \ge 1$ satisfy $n/2 <p_1 \le p_2 < n/4$. 
This is obviously an erratum. 
The correct condition is that 
$ p_1,  \,  p_2 \ge 1$ satisfy 
$ n/2 > p_1 \ge  p_2 > n/4$. 
\end{rem}

We  now  formulate our inverse scattering problem. 
\\
{\bf Inverse scattering problem: }
Given the scattering operator $S$ with the domain $H^1(\R^n)$, 
determine the interaction potential $V$. 

This paper presents a reconstruction procedure of the interaction potential $V(x)$ 
from the scattering operator defined in Theorem \ref{thm:1-1}.      

We denote  the multiplication operator  with a fixed function $V(x)$ as V,  
the Schwartz class as $\Cal{S}$, the weighted $L^2$ space as 
$L^{2,s}$ and 
the set of compactly supported smooth functions as $C_0^{\infty}$. 
The Fourier transform 
of $f$ is denoted as $\widehat{f}$ or $\Cal{F}f$.  
Let $<\cdot, \cdot>_{L^2}$ be the inner product in $L^2(\R^n)$ and 
put 
\[ \Cal{S}_0 =\{ f\in \Cal{S}(\R^n) \, ; \, \widehat{f}\in C_0^{\infty}(\R^n)\}.
\]

\begin{thm}\label{thm:1-2}
Let $3\le n \le 6$ and $2\delta >n$. 
Assume that  
$V\in \Cal{V}\cap L^{2,1+\delta}(\R^n)$ 
is a radial, non-negative and non-increasing function.    
In addition, suppose that 
$V$ is a compact operator from $L^2(\R^n)$ to $L^{2,1+\delta}(\R^n)$. 
Then for any $\varphi \in \Cal{S}_0$, we have 
\[
   \lim_{|v|\to\infty} \left< i(S-I)\Phi_v, \Phi_v \right>_{L^2} 
   =
   \int_{\R^n} \widehat{V}(\xi) G(\xi) \, d\xi < \infty, 
\]
where $\Phi_v (x) = e^{iv\cdot x}\varphi (x) $, $v\in \R^n$ and 
\[
   G(\xi)= \int_{\R} \left| \Cal{F} \left( |U_0 (t) \varphi \right|^2 ) (\xi)
   \right|^2 \, dt. 
\]

\end{thm}

As is proved in \cite{sasaki-watanabe}, 
the identity 
\[
    \lim_{\ve \to 0}\frac{1}{\ve^3}  \left< i(S-I)(\ve \varphi), \varphi  \right>_{L^2} 
     = \int_{\R^n} \widehat{V}(\xi) G(\xi) \, d\xi
\]
holds for any $\varphi \in H^1(\R^n)$. Hence we have

\begin{cor}
Let $3\le n \le 6$ and  $2\delta >n$. 
Assume that  
$V\in \Cal{V}\cap L^{2,1+\delta}(\R^n)$ 
is a radial, non-negative and non-increasing function.    
In addition, suppose that 
$V$ is a compact operator from $L^2(\R^n)$ to $L^{2,1+\delta}(\R^n)$.
Then for any 
$\varphi \in \Cal{S}_0$ and $\ve >0$, we have 
\[
   \lim_{|v|\to\infty} \left< i(S-I)\Phi_v, \Phi_v \right>_{L^2} 
   =
   \lim_{\ve \to 0}\frac{1}{\ve^3}  \left< i(S-I)(\ve \varphi), \varphi  \right>_{L^2}, 
\]
where $\Phi_v (x) = e^{iv\cdot x}\varphi (x) $, $v\in \R^n$.
\end{cor}

This corollary shows that in the case of the RH equation,  
the high-velocity  method on the inverse scattering problem is equivalent to  
the small-amplitude method on it.

Let $\varphi_{\lambda}(x)=\varphi((\lambda +1)x)$ and  put
  \begin{align*}
   G(\xi, \lambda) &:= \int_{\R} \left| \Cal{F} \left( \left| 
                                  U_0(t)\varphi_{\lambda}
                                  \right|^2 \right) (\xi) \right|^2 \, dt \\
                             &= \int_{\R} \left| 
                                   e^{-it\xi^2} 
                                    \int_{\R^n} e^{2it\xi \cdot \eta} 
                                    \widehat{\varphi_{\lambda}}(\xi -\eta)
                                      \widehat{\overline{\varphi_{\lambda}}}(\eta) \, d\eta
                                     \right|^2 \, dt   \\
                              &= \int_{\R} \left| 
                                      \int_{\R^n} e^{2it\xi \cdot \eta} \left( \frac{1}{\lambda+1}\right)^2 
                                       \widehat{\varphi}\left( \frac{\xi -\eta}{\lambda +1}\right) 
                                       \widehat{\overline{\varphi}}\left( \frac{\eta}{\lambda +1}\right) \, d\eta 
                                      \right|^2 \, dt. 
  \end{align*}

 Consider the integral equation of the first kind: 
  \begin{equation}\label{eqn:IEQ-1}
    P(\lambda) = \int_{\R^n} \widehat{V}(\xi) G(\xi, \lambda)\, d\xi.
  \end{equation}

 \begin{thm}\label{thm:1-3}
   Let $\Gamma \subset \R$ be a compact set. 
   Assume that $2\le n \le 6$. 
   Then for any $\varphi \in H^1(\R^n)$, the integral operator $T_G$:   
   \[
      (T_G f)= \int_{\R^n} f(\xi) G(\xi, \lambda) \, d\xi
   \]
 is a compact operator from $H^{k}(\R^n)$ to $L^2(\Gamma)$ 
 for $k>n/2$. 
 \end{thm}

The Picard's theorem for an equation of the first kind with a compact operator 
gives an explicit solution to the integral equation 
\eqref{eqn:IEQ-1} (see, e.g., Kress \cite[Theorem 15.18]{kress}). 
To state our theorem on the reconstruction problem,  
we give a definition of the singular system of the 
compact operator. 

\begin{defn}
Let $X$ and $Y$ be  Hilbert space, $\Cal{A}\, :\, X\to Y$ be a compact linear operator, and 
$\Cal{A}^* \, : \, Y \to X$ be its adjoint.  
Singular values of $\Cal{A}$ is 
the non-negative square roots of the eigenvalue of non-negative self-adjoint compact 
operator $\Cal{A}^* \Cal{A} \, : \, X \to X$. 
The singular system of $\Cal{A}$ is 
the system $\{ \mu_n, \varphi_n, g_n\}$, $n\in \Bbb{N}$,
where 
 $\varphi_n \in X$ and $g_n\in Y$ are orthonormal sequences such that 
$\Cal{A}\phi_n = \mu_n g_n$ and $\Cal{A}^*g_n = \mu_n \phi_n$ for all 
$n\in \Bbb{N}$. 

\end{defn}

We denote the null-space of the operator $T$ by $\Cal{N}(T)$. 

 \begin{thm}\label{thm:1-4}
Let $3\le n \le 6$ and  
$2\delta >n$. 
Assume that  
$V\in \Cal{V}\cap L^{2,1+\delta}(\R^n)$ 
is a radial, non-negative and non-increasing function.    
In addition, suppose that 
$V$ is a compact operator from $L^2(\R^n)$ to $L^{2,1+\delta}(\R^n)$. 
Then for any  $\varphi \in \Cal{S}_0$, the function 
\[
   P(\lambda):=\lim_{|v|\to\infty} 
   \left< i(S-I)\Phi_v(\cdot, \lambda), \Phi_v (\cdot, \lambda) \right>_{L^2}, 
   \quad 
   \Phi_v(x,\lambda)=e^{iv\cdot x}\varphi_{\lambda}(x) 
\] 
  is the $L^2$-function on a compact set $\Gamma \subset \R$. 
  Moreover, 
  letting $\{ \mu_n, \varphi_n, g_n\}$, $n\in \Bbb{N}$ be a singular system of 
  the compact operator $T_G$, the Fourier transform of the interaction potential is reconstructed by the formula:      
    \[
       \widehat{V}(\xi)= \sum_{n=1}^{\infty} \frac{1}{\mu_n} 
       \left< P, g_n \right>_{L^2(\Gamma)} \varphi_n
    \]
   if and only if  $P\in \Cal{N}(T^*_G)^{\perp}$ and satisfies
  
      \[
       \sum_{n=1}^{\infty} \frac{1}{\mu_n^2} \left|
         \left<P , g_n \right>_{L^2(\Gamma)} \right|^2 <\infty.
    \]
 \end{thm}

\begin{rem}
Due to the fact that $\| \widehat{f} \|_{H^k} \le C \| f \|_{L^{2,k}}$ for $k>0$, 
the Picard's theorem  can be applied to the equation \eqref{eqn:IEQ-1} 
for $V \in L^{2, 1+\delta}(\R^n)$. 
\end{rem}

 \begin{rem}\label{rem:4}
  Uniqueness theorem on the inverse scattering problem of identifying $V(x)$ holds for a bounded continuous function $V(x)$ such that 
   \[
      | V(x) | \le C| x |^{-\sigma}, \quad 2\le \sigma \le 4, \, \sigma<n
    \] 
for some $C>0$ and $\widehat{V}$ is the continuous function on $\R^n$
  (see \cite{sasaki-watanabe}). 
 \end{rem}

\subsubsection{Hartree equation} 
Consider the Hartree equation \eqref{eqn:1-1}. 
The following theorem on the scattering problem is obtained easily from 
the proof of Theorem \ref{thm:1-1} because the Hartree term $V_H(x) u_j$ has the 
same structure as in the RH equation. 
We denote a vector-valued $H^1$-function ${\bf f}=(f^{(j)})_{1\le j \le N}$ with 
$f^{(j)} \in H^1(\R^n)$ by ${\bf f} \in [H^1(\R^n)]^N$. 

\begin{thm}\label{thm:1-5}
Let $n\ge 3$. 
Assume that 
$V\in \Cal{V}$ is a radial, non-negative and non-increasing function. 
Then for any ${\bf f}_- \in [H^1(\R^n)]^N$, there exists a unique pair of functions 
${\bf f}_+ \in [ H^1(\R^n)]^N$ and $\g{ \varphi} \in [ H^1(\R^n)]^N$ such that 
\[ \| U(t)\g{\varphi} -U_0(t){\bf f}_{\pm}\|_{H^1}\longrightarrow 0, 
\qquad \text{as $t \to \pm \infty$}.
\]
In addition,  the scattering operator 
\[ S: \, [H^1(\R^n)]^N  \ni {\bf f}_- \longrightarrow {\bf f}_+\in [H^1(\R^n)]^N\]
is  homeomorphism in $[ H^1(\R^n) ]^N$.
\end{thm}

 \begin{rem}
  The scattering operator is represented as 
  \[
     (S\g{\varphi})^{(j)} (x) = \varphi^{(j)} (x) + \frac{1}{i} \int_{\R}
                                   e^{itH_0} 
                                    V_H( x, u ) u_j (x,t)\, dt, 
  \]
 $ j= 1, 2, \cdots, N$. 
 \end{rem}

We consider now a inverse scattering problem of identifying 
the interaction potential $V(x)$ from the scattering operator $S$ 
with the domain $[H^1(\R^n)]^N$.

\begin{thm}\label{thm:H-2}
Let $3\le n \le 6$ 
 and $2\delta >n$. 
Assume that  
$V\in \Cal{V}\cap L^{2,1+\delta}(\R^n)$ 
is a radial, non-negative and non-increasing function.    
In addition, suppose that 
$V$ is a compact operator from $L^2(\R^n)$ to $L^{2,1+\delta}(\R^n)$. 
Then for any  
$\varphi^{(j)}  \in \Cal{S}_0$, $j=1,2, \cdots, N$, we have 
\[
   \lim_{|v|\to\infty} \left< i\left( (S-I)\g{\Phi}_v\right)^{(j)} ,
      \Phi_v^{(j)} \right>_{L^2} 
   =
   \int_{\R^n} \widehat{V}(\xi) H^{(j)}(\xi) \, d\xi, 
\]
where $\g{\Phi}_v (x) = e^{iv\cdot x}\g{\varphi} (x) $, $v\in \R^n$ and 
\[
   H^{(j)}(\xi)= \sum_{\substack{k=1 \\ k \not=j}}^N 
                \int_{\R}  \Cal{F} \left( |U_0 (t) \varphi^{(k)} \right|^2 ) (\xi) 
                \overline{\Cal{F} \left( |U_0 (t) \varphi^{(j)} \right|^2 ) (\xi)} \, dt. 
\]

\end{thm}

As is discussed in sub-subsection \ref{subsub1-3-2},  
the high-velocity limit of the scattering operator 
is equal to the small-amplitude limit of it. 

\begin{cor}
Let $3\le n \le 6$ 
 and $2\delta >n$. 
Assume that  
$V\in \Cal{V}\cap L^{2,1+\delta}(\R^n)$ 
is a radial, non-negative and non-increasing function.    
In addition, suppose that 
$V$ is a compact operator from $L^2(\R^n)$ to $L^{2,1+\delta}(\R^n)$. 
Then for any  
$\varphi^{(j)} \in \Cal{S}_0$, $j=1,2,\cdots , N$ and $\ve >0$, we have 
\[
   \lim_{|v|\to\infty} \left< \left( i(S-I)\g{\Phi}_v\right)^{(j)} , 
    {\Phi}_v^{(j)}  \right>_{L^2} 
   =
   \lim_{\ve \to 0}\frac{1}{\ve^3}  
   \left< \left( i(S-I)(\ve \g{\varphi})\right)^{(j)} , 
    \varphi^{(j)} \right>_{L^2}, 
\]
where $\g{\Phi}_v (x) = e^{iv\cdot x}\g{\varphi} (x) $, $v\in \R^n$.
\end{cor}

Let $\g{\varphi}_{\lambda}(x)=\g{\varphi}((\lambda +1)x)$ 
and put 
  \begin{align*}
   H^{(j)}(\xi, \lambda) =  \sum_{\substack{k=1 \\ k \not=j}}^N 
                \int_{\R}  \Cal{F} \left( \left| U_0 (t) \varphi_{\lambda}^{(k)} \right|^2 \right) (\xi) 
                \overline{\Cal{F} \left( \left| U_0 (t) \varphi_{\lambda}^{(j)}  \right|^2 \right) (\xi)} \, dt.
  \end{align*}
 Consider the integral equation of the first kind: 
  \begin{equation}\label{eqn:IEQ-2}
    P^{(j)}(\lambda) = (T_H \widehat{V} )^{(j)}(\lambda) 
    :=\int_{\R^n} \widehat{V}(\xi) H^{(j)}(\xi, \lambda)\, d\xi.
  \end{equation}

 \begin{thm}\label{thm:H-3}
   Let $\Gamma \subset \R$ be a compact set. 
   Assume that $2\le n \le 6$. 
   Then for any $\g{\varphi} \in [ H^1(\R^n)]^N $,  
   the integral operator $T_H$ 
  is a compact operator from $H^{k}(\R^n)$ to $L^2(\Gamma)$ 
  for $k>n/2$. 
 \end{thm}

The same  argument as in sub-subsection \ref{subsub1-3-2} leads us a 
reconstruction formula.  

 \begin{thm}\label{thm:1-8}
Let $3\le n \le 6$ 
 and $2\delta >n$. 
Assume that  
$V\in \Cal{V}\cap L^{2,1+\delta}(\R^n)$ 
is a radial, non-negative and non-increasing function.    
In addition, suppose that 
$V$ is a compact operator from $L^2(\R^n)$ to $L^{2,1+\delta}(\R^n)$. 
Then for any  $\varphi^{(j)} \in \Cal{S}_0$, 
$j=1,2, \cdots, N$, 
the function 
\[
   P^{(j)}(\lambda):=\lim_{|v|\to\infty} 
  \left< \left( i(S-I)\g{\Phi}_v\right)^{(j)} , 
    {\Phi}_v^{(j)}  \right>_{L^2} , 
   \quad 
   \g{\Phi}_v(x,\lambda)=e^{iv\cdot x}\g{\varphi}_{\lambda}(x) 
\] 
  is the $L^2$-function on a compact set $\Gamma \subset \R$. 
  Moreover, 
  letting $\{ \mu_n, \varphi_n, g_n\}$, $n\in \Bbb{N}$ be a singular system of 
  the compact operator $T_H$, the Fourier transform of the interaction potential is reconstructed by the formula:      
    \[
       \widehat{V}(\xi)= \sum_{n=1}^{\infty} \frac{1}{\mu_n} 
       \left< P^{(j)}, g_n \right>_{L^2(\Gamma)} \varphi_n
    \]
   if and only if  $P^{(j)} \in \Cal{N}(T^*_H)^{\perp}$ and satisfies
  
      \[
       \sum_{n=1}^{\infty} \frac{1}{\mu_n^2} \left|
         \left<P^{(j)} , g_n \right>_{L^2(\Gamma)} \right|^2 <\infty.
    \]
 \end{thm}

\subsubsection{Hartree-Fock equation} 
Consider the HF equation \eqref{eqn:1-2}. 
In contrast to H equation, the large data scattering  
for HF equation in $H^1(\R^n)$ space 
dose not follow in the same way as in the proof of 
Theorem \ref{thm:1-1}. 
It still remains to be a poorly understood problem, 
although basic results---the global existence and the $L^2$-conservation 
law of solutions---was obtained by Isozaki \cite{isozaki1983}. 
We here state results on the small data scattering in the 
space $H^1(\R^n)$ and the large data scattering in a weighted space 
because of a difference for assumptions on $V$.

The following theorem on the small data scattering follows from 
the result in Mochizuki \cite{mochizuki}. 

 \begin{thm}\label{thm:HF-1}
  Assume that 
  $V(x)$ satisfies  
   \[ |V(x) | \le C_V |x|^{-\sigma}, \qquad 
                  \text{$2\le \sigma \le n$ and $\sigma <n$} 
   \]
   for some $C_V>0$. 
Then for any ${\bf f}_- \in [ H^1(\R^n)]^N$, there exists a unique pair of functions 
${\bf f}_+ \in [  H^1(\R^n)]^N$ and $\g{\varphi} \in [ H^1(\R^n)]^N$ such that 
\[ \| U(t)\g{\varphi} -U_0(t){\bf f}_{\pm}\|_{H^1}\longrightarrow 0, 
\qquad \text{as $t \to \pm \infty$}.
\]
In addition,  the scattering operator 
 $
    S: \,  {\bf f}_- \longrightarrow {\bf f}_+
 $
is defined on 
$[ H^1_{\ve}(\R^n)]^N
  =\{ \g{\varphi }\in [H^1(\R^n)]^N;  \|\varphi^{(j)}\|_{H^1} <\ve, j=1,\cdots, N \}
$ for 
small $\ve>0$ depending on $C_V$.
\end{thm}

The large data scattering is stated as follows:   

 \begin{thm}[Wada \cite{wada}]\label{thm:HF-1-1}
   Let $\ell, m \in \Bbb{N}$. Assume that 
   $V(x)$ satisfies  
    \[ |V(x) | \le C |x|^{-\sigma}, \qquad 
                  \text{$\frac{4}{3} < \sigma < \min (4, n)$} 
    \]
    for some $C>0$.  
    Suppose that $m \ge 2$ if $\sigma \le \sqrt{2}$. 
 Then for any ${\bf f}_- \in \sum^{\ell, m}$, there exists a unique pair of functions 
${\bf f}_+ \in   \sum^{\ell, m}$ and $\g{\varphi} \in \sum^{\ell,m}$ such that 
\[ \| U(t)\g{\varphi} -U_0(t){\bf f}_{\pm}\|_{\sum^{\ell,m}}\longrightarrow 0, 
\qquad \text{as $t \to \pm \infty$}.
\]
In addition,  the scattering operator 
 $
    S: \,  {\bf f}_- \longrightarrow {\bf f}_+
 $
is defined on 
   \[
       \Sigma^{\ell,m} =
          \{ \g{\varphi} \in [ L^2(\R^n)]^N \, ;  \, \|\g{\varphi}\|_{\Sigma^{\ell,m}}^2
          =\sum_{|\alpha| \le \ell}  \|  \nabla^{\alpha}\g{\varphi} \|^2_{L^2} + 
           \sum_{ | \beta | \le m} \|  x^{\beta}\g{\varphi}  \|^2_{L^2} < \infty \}.
  \]
 \end{thm}

 \begin{rem}
  Due to the fact that in each case, the scattering operator is represented as  
  \begin{align*}
     (S \g{\varphi})^{(j)} (x) &= \varphi^{(j)} (x) + \frac{1}{i} \int_{\R}
                                   e^{itH_0} 
                                   (F_{HF}({\bf u}))^{(j)}
                                     \, dt, \\
    (F_{HF}({\bf u}))^{(j)} &= 
                                   V_H(x)u_j (x,t)+ \int_{\R^n} V_F (x,y) u_j (y,t)\, dy, 
  \end{align*}
 $ j= 1, 2, \cdots, N$, our reconstruction formula given below 
 is valid for  both of the scattering although assumptions on $V$ are different. 
 \end{rem}

We consider now a inverse scattering problem of identifying 
the interaction potential $V(x)$ from the scattering operator $S$ 
with the domain $[ H^1_{\ve}(\R^n)]^N$ or $\sum^{\ell, m}$.

\begin{thm}\label{thm:HF-2}
Let $2\le n \le 6$.  Assume that  
$V$ satisfies the assumption in Theorem \ref{thm:HF-1} or 
Theorem \ref{thm:HF-1-1}.    
In addition, suppose that 
$V$ is a compact operator from $L^2(\R^n)$ to $L^{2,1+\delta}(\R^n)$ 
 with $2\delta >n$. 
Then for any $\varphi^{(j)}  \in \Cal{S}_0$, $j=1,2, \cdots, N$, we have 
\[
   \lim_{|v|\to\infty} \left< i\left( (S-I) {\bf \Phi}_v\right)^{(j)} ,  \Phi_v^{(j)} \right>_{L^2} 
   =
   \int_{\R^n} \widehat{V}(\xi) H^{(j)}_{HF}(\xi) \, d\xi, 
\]
where  $\Phi_v^{(j)} (x) = e^{iv\cdot x}\varphi^{(j)} (x) $, $v\in \R^n$ and 
\begin{align*}
   H^{(j)}_{HF}(\xi) &= \sum_{k=1}^N 
                \int_{\R}  \Cal{F} \left( |U_0 (t) \varphi^{(k)} \right|^2 ) (\xi) 
                \overline{\Cal{F} \left( |U_0 (t) \varphi^{(j)} \right|^2 ) (\xi)} \, dt \\
             &\hspace{2em}- \sum_{k=1}^N \int_{\R}  \left| 
                 \Cal{F} \left(  \left(U_0(t) \varphi^{(j)} \right) 
                \overline{ \left( U_0(t) \varphi^{(k)} \right) } \right) (\xi) \right|^2
                 \, dt. 
\end{align*}

\end{thm}
Similarly to the RH equation and H equation, the high-velocity limit of the 
scattering operator is 
equal to the small-amplitude limit of it. 

\begin{cor}
Let $2\le n \le 6$. 
 Assume that  
$V$ satisfies the assumption in Theorem \ref{thm:HF-1} or 
Theorem \ref{thm:HF-1-1}.    
In addition, suppose that 
$V$ is a compact operator from $L^2(\R^n)$ to $L^{2,1+\delta}(\R^n)$ 
 with $2\delta >n$. 
Then for any 
$\varphi^{(j)} \in \Cal{S}_0$, $j=1,2, \cdots , N$ and $\ve >0$, we have 
\[
   \lim_{|v|\to\infty} \left< \left( i(S-I){\bf \Phi}_v\right)^{(j)} ,  \Phi_v^{(j)}  \right>_{L^2} 
   =
   \lim_{\ve \to 0}\frac{1}{\ve^3}  
   \left< \left( i(S-I)(\ve \g{\varphi})\right)^{(j)} , \varphi^{(j)}  \right>_{L^2}, 
\]
where $\Phi_v^{(j)} (x) = e^{iv\cdot x}\varphi^{(j)} (x) $, $v\in \R^n$.
\end{cor}

Let $\g{\varphi}_{\lambda}(x)=\g{\varphi}((\lambda +1)x)$ and 
  \begin{align*}
   H^{(j)}_{HF}(\xi, \lambda) &= \sum_{k=1}^N 
                \int_{\R}  \Cal{F} \left(  \left| U_0 (t) \varphi_{\lambda}^{(k)} \right|^2 \right) (\xi) 
                \overline{\Cal{F} \left( \left| U_0 (t) \varphi_{\lambda}^{(j)} \right|^2 \right) (\xi)} \, dt \\
             &\hspace{2em}- \sum_{k=1}^N \int_{\R}  \left| 
                 \Cal{F} \left(  \left(U_0(t) \varphi_{\lambda}^{(j)} \right) 
                \overline{ \left( U_0(t) \varphi_{\lambda}^{(k)} \right) } \right) (\xi) \right|^2
                 \, dt. 
\end{align*}

 Consider the integral equation of the first kind: 
  \begin{equation}\label{eqn:IEQ-3}
    P^{(j)}(\lambda) = (T_{HF} \widehat{V} )^{(j)}(\lambda) 
    :=\int_{\R^n} \widehat{V}(\xi) H^{(j)}_{HF}(\xi, \lambda)\, d\xi.
  \end{equation}

 \begin{thm}\label{thm:HF-3}
   Let $\Gamma \subset \R$ be a compact set. 
   Assume that $2\le n \le 6$. 
   Then for any $\varphi^{(j)} \in H^1(\R^n)$, $j=1,\cdots, N$,  
   the integral operator $T_{HF}$ 
 is a compact operator from $H^{k}(\R^n)$ to $L^2(\Gamma)$ 
 for $k>n/2$. 
 \end{thm}

Similarly to the RH equation and the H equation, the Picard's theorem allows us to obtain a 
reconstruction formula of $\widehat{V}$ in term of the singular system of $T_{HF}$.

 \begin{thm}
  Let $2\le n \le 6$. 
  Assume that  
$V$ satisfies the assumption in Theorem \ref{thm:HF-1} or 
Theorem \ref{thm:HF-1-1}.    
In addition, suppose that 
$V$ is a compact operator from $L^2(\R^n)$ to $L^{2,1+\delta}(\R^n)$ 
 with $2\delta >n$. 
Then for any  $\g{\varphi} \in  [\Cal{S}_0]^N$, the function 
\begin{align*}
   P^{(j)} (\lambda)
   &:=\lim_{|v|\to\infty} 
   \left< i\left( (S-I)\g{\Phi}_v(\cdot, \lambda)\right)^{(j)}, 
    \Phi_v^{(j)} (\cdot, \lambda) \right>_{L^2}, \quad j=1, \cdots N, \\
   \quad 
   \g{\Phi}_v(x,\lambda)
   &=e^{iv\cdot x} \g{\varphi} ((\lambda+1)x) 
\end{align*} 
  is the $L^2$-function on a compact set $\Gamma \subset \R$. 
  Moreover, 
  letting $\{ \mu_n, \varphi_n, g_n\}$, $n\in \Bbb{N}$ be a singular system of 
  $T_{HG}$, the Fourier transform of the interaction potential is reconstructed by the formula:      
    \[
       \hat{V}(\xi)= \sum_{n=1}^{\infty} \frac{1}{\mu_n} 
       \left< P^{(j)}, g_n \right>_{L^2(\Gamma)} \varphi_n
    \]
   if and only if  $P^{(j)}\in \Cal{N}(T^*_{HF})^{\perp}$ and satisfies
      \[
       \sum_{n=1}^{\infty} \frac{1}{\mu_n^2} \left|
         \left<P^{(j)} , g_n \right>_{L^2(\Gamma)} \right|^2 <\infty.
    \]
 \end{thm}
 
\begin{thm}\label{thm:uniqueness}
Let $2\le n \le 6$. 
Assume that $V_{\sharp}, \sharp = 1,2$ satisfy the assumption in 
Theorem \ref{thm:HF-2}. 
Let $S_{\sharp}$ are the scattering operator for the HF equation 
\eqref{eqn:1-2} with interaction potential $V_{\sharp}$. 
If $S_1=S_2$, then we have $V_1=V_2$. 
\end{thm}

The structure of this paper is as follows. 
Section \ref{sec:2} is devoted to an analysis of high-velocity analysis 
of the scattering operator.  
A time-space estimate on $V_H(x,u) U_0(t)\Phi_v$ plays a important role. 
We give proofs of 
Theorem \ref{thm:1-3}, Theorem \ref{thm:H-3} and Theorem \ref{thm:HF-3} in Section \ref{sec:3}. 
It will be shown that 
the set of functions $\{T_G f \}$, $\{T_H f \}$ and $\{T_{HF} f \}$ 
are equicontinuous and equibounded in 
the set of continuous functions $C(\Gamma)$. 
Section \ref{sec:4} gives a proof of Theorem \ref{thm:uniqueness}. 


\section{High velocity limit of the scattering operator}\label{sec:2}
In this section, we analyze the asymptotic behavior of the scattering operator 
for the  RH equation, the H equation and the HF equations. 
Due to the similarity of  the proof, we give a proof in detail 
only for the case of the RH equation.  

Consider the RH equation \eqref{restricted Hartree}. 
Let $S$ be the scattering operator for  \eqref{restricted Hartree} 
defined in Theorem \ref{thm:1-1}.
Our goal in this section is to prove the following theorem.
 
 \begin{thm}\label{thm:2-1}
Let $3\le n \le 6$ 
and $2\delta >n$. 
Assume that  
$V\in \Cal{V}\cap L^{2,1+\delta}(\R^n)$ 
is a radial, non-negative and non-increasing function.    
In addition, suppose that 
$V$ is a compact operator from $L^2(\R^n)$ to $L^{2,1+\delta}(\R^n)$. 
Then for any
$\varphi \in \Cal{S}_0$, we have 
\[
    \left< i(S-I)\Phi_v, \Phi_v \right>_{L^2} 
   =
   \int_{\R^n} \widehat{V}(\xi) G(\xi) \, d\xi 
   + R(v), 
\]
where  $\Phi_v (x) = e^{iv\cdot x}\varphi (x) $, $v\in \R^n$ and 
 \begin{align*}
   G(\xi)&= \int_{\R} \left| \Cal{F} \left( |U_0 (t) \varphi \right|^2 ) (\xi)
   \right|^2 \, dt, \\
    R(v) &= O(|v|^{-2}), \quad |v|\to \infty.
 \end{align*}

\end{thm}

\subsection{Preliminary lemmas}
In order to prove Theorem \ref{thm:2-1}, we need some 
lemmas. 

\begin{lem}\label{lem:Enss-Weder}
Let $n\ge 2$ and $s>1$. 
Assume that 
$q$ is a compact operator from $L^2(\R^n)$ to $L^{2,s}(\R^n)$.  
Then for any $\varphi \in \Cal{S}_0$, 
there exist a positive constant
$C$ such that 
\[ \int_{-\infty}^{\infty} 
 \| q U_0(t)e^{iv\cdot x}\varphi\|_{L^2} \, dt \le 
 \frac{C}{|v|}
\]
for $|v|$ large enough. 
\end{lem}
{\it Proof. }
The proof will be found in \cite[Lemma 2.2]{Enss-Weder1995} 
and its proof. 
\hfill $\Box$

We establish a similar estimate for the  RH equation.


\begin{lem}\label{lem:RH2-1}
Let $n\ge 3$ and $s>1$. Assume that 
$V\in\Cal{V}$ is a radial, non-negative and non-increasing function. 
Suppose that 
$V$ is a compact operator from $L^2(\R^n)$ to $L^{2,s}(\R^n)$. 
Then for any $\varphi \in \Cal{S}_0$, 
there exist a positive constant
$C$  such that 
\[ \int_{-\infty}^{\infty} 
 \|  V_H U_0(t)\Phi_v \|_{L^2} \, dt \le 
 \frac{C}{|v|}
\]
for $|v|$ large enough, 
where $V_H(x, u)= (V*|u(\cdot,t)|^2)(x)$ and 
$u(t)=u(x,t)$ is the scattering solution for the RH equation \eqref{restricted Hartree}.  

\end{lem}

 {\it Proof. }
Due to the $L^2$ boundedness of the scattering solution to the RH equation 
\eqref{restricted Hartree}, we have  
 \[
    \left| \Cal{F}(| u |^2)(t, \xi) \right|  
     \le \| |u (t)|^2  \|_{L^1} \le \| f_- \|_{L^2}^2.
 \]
From this inequality and the identity $\widehat{V_H} = \widehat{V} \Cal{F}(|u|^2)$,  
one has 
  \begin{align*}
      \|  V_H(\cdot, u) U_0(t) \Phi_v \|_{L^2} 
     &=
      \| (\widehat{V} \widehat{|u|^2})* \Cal{F}(U_0(t)\Phi_v) \|_{L^2} \\
     &\le 
     \sup_{t,\xi} \left| \widehat{|u|^2}(t,\xi) \right|  \|  \widehat{V} * \Cal{F}(U_0(t)\Phi_v) \|_{L^2} \\
     &\le 
      C \| V U_0(t) \Phi_v \|_{L^2}.
  \end{align*}
Then, applying Lemma \ref{lem:Enss-Weder} to the right hand side of the above 
inequality achieves the desired estimate. 
 \hfill $\Box$

Let $u$ be the scattering solution to \eqref{restricted Hartree}.
Consider the wave operator 
$\Omega_- \, : \, H^1 \ni f_-\to u(0)\in H^1$.


\begin{lem}\label{lem:RH2-2}
Let $n\ge 3$ and $\Phi_v=e^{iv\cdot x}\varphi$. 
Assume that $V(x)$ satisfies the same condition as in Lemma \ref{lem:RH2-1}. 
Then for any $\varphi \in \Cal{S}_0$,  
we have  
\[ \| (\Omega_- -I)U_0(t)\Phi_v\|_{L^2}=O(|v|^{-1})\]
as $|v|\to \infty$ uniformly in $t\in \R$.  
\end{lem}

{\it Proof. }
In view of  the representation of the wave operator, one has 
\[\left<\Omega_- f, g\right>_{L^2}-\left<f,g\right>_{L^2}=
 i \int_{-\infty}^0 \left< 
 V_H(x, u)u(s), U_0(s)g\right>_{L^2}\, ds. 
\]
Then, Lemma \ref{lem:RH2-1} and 
 the duality argument enables us to obtain 
\begin{align*}
 \| (\Omega_- - I)U_0(t)\Phi_v\|_{L^2} &\le 
  \int_{-\infty}^{\infty}  \| u(s) \|_{L^2}  \| V_H(\cdot, u) U_0(s)g\|_{L^2}\, ds \\
  &\le \frac{C}{|v|}.
\end{align*}
Here $C$ is  a positive constant independent of $t$. 
This completes the proof.
\hfill $\Box$

 \begin{lem}\label{lem:RH2-3}
Let $n\ge 2$, $\delta>0$ and $\Phi_v=e^{iv\cdot x}\varphi$, $v\in \R^n$. 
Assume that $V$ is a compact operator from $L^2(\R^n)$ to $L^{2, 1+\delta}(\R^n)$. 
 Then for any $\varphi \in \Cal{S}_0$ and $y\in \R^n$, there exist positive constants 
 $C_1$, $C_2$ and $C_3$ such that  
  \begin{align*}
    \| V(\cdot -y) U_0(t) \Phi_v \|_{L^2} 
    &\le 
     C_1 \left( 1+ \frac{ |vt|}{ 4 } \right)^{-3} 
      + C_2 ( 1+ | vt | )^{-3/2} \\
     & + C_3 \left\{ 1+ \left( \frac{3}{8} | vt | - | y | \right)^2 \right\}^{-(1+\delta)/2}.
  \end{align*}
 \end{lem}

{\it Proof. }
The proof of this lemma is almost the same as in \cite{Enss-Weder1995}. 
\hfill $\Box$

We denote by $L^{p}(\R ; L^q)$ the set of $L^q$-valued $L^p$ functions. 
\begin{lem}\label{lem:RH2-4}
 Let $2\le n \le 6$. Then for any $\varphi \in H^1(\R^n)$, 
 \[
    \| U_0(\cdot) \varphi \|_{L^4(\R; L^4)} \le C \| \varphi \|_{H^1}, \qquad C>0.
 \]
\end{lem}

{\it Proof. }
The proof will be found in \cite[Proposition 6]{sasaki-watanabe}.
\hfill $\Box$

\subsection{Proof of Theorem \ref{thm:2-1}}
We are now in a position to prove Theorem \ref{thm:2-1}. 
Let $F_{RH}(u)= (V*|u(t)|^2)u(t)$. 
We break the scattering operator in four parts: 
 \begin{equation*}
  \left< i(S-I)\Phi_v, \Phi_v \right>_{L^2}
   =L(v)+R_1(v)+R_2(v)+R_3(v), 
 \end{equation*} 
where 
 \begin{align*}
   L(v) &=  \int_{-\infty}^{\infty}\left< 
                 F_{RH}( U_0(s)\Phi_v ),  U_0(s)\Phi_v\right>_{L^2}\, ds, \\
   R_1(v)&=
    \int_{-\infty}^{\infty}\left< 
     \left[ V*\left\{ u(s)-U_0(s)\Phi_v\right\}\overline{U_0(s) \Phi_v} \right] U_0(s)\Phi_v, 
     U_0(s)\Phi_v \right>_{L^2} \, ds, \\
  R_2(v)&=
   \int_{-\infty}^{\infty}\left< 
     \left[ V*u(s) \left\{ \overline{u(s)-U_0(s)\Phi_v}\right\} \right] U_0(s)\Phi_v, 
     U_0(s)\Phi_v \right>_{L^2}\, ds, \\
  R_3(v)&=
   \int_{-\infty}^{\infty}\left< 
     \left\{ V* |u(s)|^2\right\} \left\{ u(s)- U_0(s)\Phi_v\right\} , 
     U_0(s)\Phi_v \right>_{L^2}\, ds. 
 \end{align*}
Here $u(t)$ is the scattering solution to \eqref{restricted Hartree}.

To calculate the leading term $L(v)$, 
we first observe that the identity 
 \[ 
    \left< F_{RH} (U_0(t)\Phi_v), U_0(t)\Phi_v \right>_{L^2} 
     = \left< F_{RH}(U_0(t) \varphi) , U_0(t) \varphi \right>_{L^2}
 \]
holds.  In fact, by  using 
the identity 
\[ (U_0(s)\Phi_v)(x)=e^{i( v\cdot x - |v|^2s/2)}
 (U_0(s)\varphi)(x-vs),
\]
and the change of variables 
 $s=t/|v|$,  $x'=x-vt$ and $y'=y-vt$, 
 we obtain 
 \begin{align*}
  & \left< F_{RH} (U_0(t)\Phi_v),  U_0(t)\Phi_v \right>_{L^2} \\ 
  &= 
  \int_{\R^n}  \left| e^{i(v\cdot x-|v|^2 t/2)}(U_0(t)\varphi)(x-vt)  \right|^2 
  \left(  V * | (U_0(t)\varphi ) (\cdot- vt) |^2 \right)(x) \, dx \\
  &= 
  \int_{\R^n}  \left| (U_0(t)\varphi)(x') \right|^2 
   \left( \int_{\R^n} V(x'+vt-(y'+vt)) | U_0(t) \varphi(y') |^2 \, dy' \right) \, dx \\
  &=
  \int_{\R^n}  V_H( x, U_0(t)\varphi) (U_0(t)\varphi)(x') \overline{(U_0(t)\varphi)(x')} 
    \, dx' \\ 
  &=
   \left< F_{RH}(U_0(t)\varphi), U_0(t)\varphi \right>_{L^2}. 
 \end{align*}
The Plancherel's theorem implies that  
  \[ 
     \left< F_{RH}(U_0\varphi), U_0(t)\varphi \right>_{L^2} 
      =\left< \widehat{V} \Cal{F}(|U_0(t)\varphi |^2), \Cal{F}(|U_0(t)\varphi |^2)\right>_{L^2}. 
  \]
Then, the Fubini's thoerem yields the expression of the leading term 
  \[
     L(v) = \int_{\R^n} \widehat {V}(\xi) \left( \int_{\R} 
               \left| \Cal{F}( | U_0 (t) \varphi |^2 ) \right|^2 \, dt \right) \, d\xi.
  \]  
Here we note that $L(v)$ is bounded.  In fact, 
thanks to  Lemma \ref{lem:RH2-4}, one gets  
 \begin{align*}
    | L(v) | 
    & \le 
     \| \widehat{V} \|_{L^{\infty}} \int_{\R} \int_{\R^n}
     \left| \Cal{F}( | U_0 (t) \varphi |^2 ) \right|^2 \, d\xi dt \\
    & \le 
    C \| V \|_{L^{1}}
    \int \left\| \Cal{F}( | U_0 (t) \varphi |^2 ) \right\|_{L^2}^2 \, dt \\
    & =
    C \|  U_0(t) \varphi \|_{L^4( \R ; L^4)}^4 \\
    & \le 
     C \| \varphi \|_{H^1}^4. 
  \end{align*}


Next, we will show that $R_3(v)=O(|v|^{-2})$ as $|v|\to \infty$. 
Thanks to   
Lemma \ref{lem:RH2-1} and Lemma \ref{lem:RH2-2}, one has   
 \begin{align*}
  \left|  R_3(v) \right| & \le 
   \int_{-\infty}^{\infty} \left| \left< 
     (\Omega_-  - I)( U_0(s) \Phi_v ), 
     V_H (\cdot, u) U_0 (s) \Phi_v \right>_{L^2} \right| \, ds \\ 
   & \le 
     \int_{\R} \| (\Omega_- -I) ( U_0(s) \Phi_v) \|_{L^2} \, 
     \| V_H (\cdot, u) ( U_0(s) \Phi_v) \|_{L^2}\, ds  \\
   & \le 
     \frac{C}{ | v | } 
     \int_{\R} \| V_H(\cdot, u) ( U_0(s) \Phi_v) \|_{L^2}\, ds \\
   & \le 
     \frac{C}{ | v |^2 }, 
 \end{align*}
due to the fact that $u(t)= \Omega_{-} \left( U_0(t) \phi \right)$. 

We will claim that 
$R_j(v)=O(|v|^{-2})$, $j=1,2$ as $|v|\to \infty$. 
Due to the fact that  
 \begin{align*}
    \| (V* ab) c\|_{L^2}^2 
    & \le
     \| a\|_{L^2}^2 \int_{\R^n } \, | b(y) |^2 \left( 
      \int_{\R^n} | V(x-y) c(x) |^2 \, dx  \right) \, dy, 
 \end{align*}
we obtain 
 \begin{align*}
  \left| R_1(v)  \right| & \le 
    \int_{\R} \| \left( V* \{ (\Omega_- - I) ( U_0(s) \Phi_v )\} \overline{ U_0 (s) \Phi_v} \right) 
     U_0(s) \Phi_v \|_{L^2} \, \| U_0(s) \Phi_v \|_{L^2} \, dt \\
  & \le 
    \| \varphi \|_{L^2}
    \int_{\R} \| \left( V* \{ (\Omega_- - I) ( U_0(s) \Phi_v )\} \overline{ U_0 (s) \Phi_v} \right) 
     U_0(s) \Phi_v \|_{L^2}\, dt \\
  & \le 
    \| \varphi \|_{L^2} \int_{\R} \| (\Omega_- -I) U_0 (s) \Phi_v \|_{L^2} \\
   &\hspace{1em} 
   \left\{
   \int_{\R^n}   | \overline{ (U_0(s) \Phi_v) (y)} |^2
     \left( \int_{\R^n} \left| V(x - y) ( U_0(s) \Phi_v)(x) \right|^2\, dx \right) \, dy \right\}^{1/2}\,  dt. 
 \end{align*}
Thanks to 
Lemma \ref{lem:RH2-2} and Lemma \ref{lem:RH2-3}, one has
 \begin{align*}
  | R_1 (v) | 
   & \le 
   \frac{C}{ | v | } \left( R_1^{(1)}(v) + R_1^{(2)}(v)  \right), 
 \end{align*}
where
 \begin{align*}
  R_1^{(1)}( v ) & = \int_{\R} \left\{ \left( 1+\frac{ | vt | }{ 4 } \right)^{-3} + 
   (1+ | vt |)^{-3/2} \right\} 
   \| U_0 ( s ) \Phi_v \|_{L^2} \, dt     \\
   R_1^{(2)}(v) &= 
    \int_{\R}\left\{ \int_{\R^n} \left( 1+ \left( \frac{3}{8} | vt | - | y | \right)^2 \right)^{-(1+\delta)}\, 
     | \overline{ U_0(t) \Phi_v } |^2 \, dy \right\}^{1/2}\, dt.
 \end{align*}
Let us prove that 
$R_1^{(1)}(v) \le C|v|^{-1}$ for some $C>0$. 
Thanks to  Lemma \ref{lem:RH2-2}, it is easy to verify that  
 \begin{align*}
  R_1^{(1)}( v ) & = \| \varphi \|_{L^2}
   \int_{\R} \left\{ \left( 1+\frac{ | vt | }{ 4 } \right)^{-3} + 
   (1+ | vt |)^{-3/2} \right\} \, dt \\
   & \le 
   \frac{C}{ | v |}.
 \end{align*}

We will show that 
$R_1^{(2)}(v)\le C|v|^{-1}$ for some $C>0$. 
By using 
Lemma \ref{lem:RH2-3} and  estimate 
$ \| U_0(t) \Phi_v \|_{L^{\infty}} \le \| \widehat{\varphi} \|_{L^1}$, 
one gets
 \begin{align*}
  R_1^{(2)}(v) & \le 
  C \| \widehat{\phi} \|_{L^1} 
   \int_{\R} \left\| \left( 1+ \left( \frac{3}{8} | vt |- | y | \right)^2 
   \right)^{-(1+\delta)/2}\right\|_{L^2}\, dt. 
 \end{align*}
The polar coordinate yields that  
 \begin{align*}
  \left\| \left( 1+ \left( \frac{3}{8} | vt |- | y | \right)^2 \right)^{-(1+\delta)/2}\right\|_{L^2}^2 
  = | \Bbb{S}^{n-1} | \int_0^{\infty} 
   \frac{ r^{n-1} }{ (1+ ( 3 | vt |/8 - r )^2 )^{(1+\delta)} } \, dr,
 \end{align*}
where $ | \Bbb{S}^{n-1} |$ is the surface area of the unit 
$(n-1)$-sphere in $n$-dimensional Euclidean space. 
It is easy to verify  that 
 \begin{align*}
  \int_0^{3 | vt | /16 } 
  \frac{ r^{n-1} }{ (1+ ( 3 | vt |/8 - r )^2 )^{(1+\delta)} }\, dr 
  & \le 
  \int_0^{ 3 | vt | /16 } 
  \frac{ r^{n-1} }{ ( 1+ ( 3 | vt | /16)^2 /4)^{(1+\delta)} }\, dr \\
  &= 
  \frac{1}{n} \frac{ (3 | vt | /16 )^n }{ (1+ ( 3 | vt |/8 )^2 /4)^{(1+\delta)}}
 \end{align*}
and 
 \begin{align*}
  \int_{3 | vt | /16 }^{\infty} 
   \frac{ r^{n-1} }{ (1+ ( 3 | vt |/8 - r )^2 )^{(1+\delta)} }\, dr
  & \le 
  2 \int_0^{ 3 | vt | /16 } 
  \frac{ (1+r)^{n-1} }{ (1+ r)^{2(1+\delta)} }\, dr \\
  &= 
  \frac{ 2}{ 2(1+\delta) -n } \left( 1+ \frac{3}{16} | vt | \right)^{n-2(1+\delta)}
 \end{align*}
for $n< 2( 1+\delta)$.  
Then we have 
  \begin{align*}
  \left\| \left( 1+ \left( \frac{3}{8} | vt |- | y | \right)^2 \right)^{-(1+\delta)/2}\right\|_{L^2}^2 
  & \le  
     \frac{C_1 (3 | vt | /16 )^n }{ (1+ ( 3 | vt |/8 )^2 /4)^{(1+\delta)}} \\
  & +
    \frac{ C_2}{ 2(1+\delta) -n } \left( 1+ \frac{3}{16} | vt |  \right)^{n-2(1+\delta)} \\
  & \le 
  C_3 \left( 1 + \frac{3}{16} | vt |\right)^{n-2(1+\delta)}
 \end{align*}
for $n<2(1+\delta)$, where $C_3$ is a positive constant depending only on $n$ and $\delta$. 
Therefore,  one has  
 \begin{align*}
  R_1^{(2)}(v) & \le 
   \frac{C}{ | v | } \int_{\R} (1+s/2)^{(n-2(1+\delta))/2}\, dx \\
 & = \frac{ C } { | v |}
 \end{align*}
 for $ n/2 < \delta$, due to  the change of  variables $3 |vt|/8=s$. 
 
 We now conclude that $ | R_1(v) | \le C/ | v |^{2}$. 
Similarly,  the remainder term $R_2(v)$ is estimated as
 \[
    |R_2(v) | \le \frac{C}{ | v |^{2} }, \qquad \text{ for $n/2 < \delta$}.
 \] 
Consequently, letting $R(v)=R_1(v)+R_2(v)+ R_3(v)$, we obtain  
 \[
    \left|   R(v) \right| 
     \le \frac{C_1}{ | v |^2 } + \frac{C_2}{ | v |^2 },
 \]
 which proves the theorem. 
 \section{Integral equations}\label{sec:3}
 In this section, we will show that the integral operators 
 $T_G$, $T_{H}$ and $T_{HF}$ are compact operators. 
 After giving a proof in detail for $T_G$ in subsection \ref{sub3-1}, 
 we give a sketch of proofs for $T_H$ and $T_{HF}$ in 
 subsection \ref{sub3-2} and \ref{sub3-3}, respectively. 
 
\subsection{Integral operator $T_G$}\label{sub3-1}
Consider the integral operator $T_G$: 
 \[ 
    (T_G f)(\lambda) = \int_{\R^n} f(\xi) G(\xi, \lambda)\, d \xi,
 \]
where 
 \[
    G(\xi, \lambda)= \int_{\R} \left|  \Cal{F} \left( | U_0(t) \varphi_{\lambda} |^2 \right) (\xi)
    \right|^2 \, dt
 \]
and $\varphi_{\lambda}(x)= \varphi ( (\lambda+1)x)$. 

Due to the Sobolev embedding theorem $H^k(\R^n) \hookrightarrow L^{\infty}(\R^n)$ 
for $2k >n$, in order to prove Theorem \ref{thm:1-3}, it suffices to verify 
that $T_G$ is a compact operator from 
$L^{\infty}(\R^n)$ to $C(\Gamma)$. 


 \begin{thm}\label{thm:3-1}
   Let $\Gamma \subset \R$ be a compact set. 
   Assume that $2\le n \le 6$. 
   Then for any $\varphi \in H^1(\R^n)$ the integral operator $T_G$ 
 is a compact operator from $L^{\infty}(\R^n)$ to $C(\Gamma)$. 
 \end{thm}

 {\it Proof. }
  Due to the fact that 
  $| |u|^2- | v |^2 | \le | u-v |^2 + 2( | u |  + | v |) | u-v |$, we have
   \begin{align*}
     | (T_G f )(\lambda)- (T_G f )(\lambda') | 
      & \le 
      T_G^{(1)}(\lambda, \lambda') + 2T_G^{(2)}(\lambda, \lambda'),
   \end{align*}
  where
   \begin{align*}
    T_G^{(1)}(\lambda, \lambda') &= 
     \int_{\R^n} | f(\xi) | \left( 
      \int_{\R} \left| \Cal{F}( |U_0(t)\varphi_{\lambda} |^2 ) (\xi)- 
       \Cal{F}( |U_0(t) \varphi_{\lambda'}|^2) (\xi) \right|^2 \, dt \right) \, d\xi , \\
   T_G^{(2)}(\lambda, \lambda') &= 
     \int_{\R^n} | f(\xi) | \left( 
      \int_{\R} \left\{
      |\Cal{F}( | U_0(t) \varphi_{\lambda} |^2)(\xi) | + | \Cal{F}( | U_0(t) \varphi_{\lambda'} |^2) (\xi) | 
      \right\} \right. \\
       & \hspace{2em} \left. \left| \Cal{F}( |U_0(t)\varphi_{\lambda} |^2 ) (\xi)- 
       \Cal{F}( |U_0(t) \varphi_{\lambda'}|^2) (\xi) \right|  \, dt \right) \, d\xi. 
   \end{align*}
Thanks to Lemma \ref{lem:RH2-4}, one has
 \begin{align*}
   T_G^{(1)}(\lambda, \lambda') 
    & \le 
    \| f \|_{L^{\infty}} \int_{\R} 
     \left\| \Cal{F} ( | U_0(t)\varphi_{\lambda} |^2 - | U_0(t)\varphi_{\lambda'} |^2 ) 
     \right\|_{L^2}^2\, dt  \\ 
    & \le 
     \| f \|_{L^{\infty}} \int_{\R} 
      \left\| ( | U_0(t)\varphi_{\lambda} | + | U_0(t)\varphi_{\lambda'} | ) 
      ( | U_0(t)\varphi_{\lambda} | - | U_0(t)\varphi_{\lambda'} | ) \right\|^2_{L^2}\, dt \\
    & \le 
     \| f \|_{L^{\infty}}\int_{\R} 
      \left\| \, | U_0(t)\varphi_{\lambda} | + | U_0(t)\varphi_{\lambda'} | \right\|^2_{L^4} \, 
      \left\| \, | U_0(t)\varphi_{\lambda} | -  | U_0(t)\varphi_{\lambda'} | \right\|^2_{L^4}\, dt\\ 
    & \le 
     \| f \|_{L^{\infty}} \left( 
      \| U_0(\cdot) \varphi_{\lambda} \|_{L^4(\R; L^4)}^2 
       + \| U_0(\cdot) \varphi_{\lambda'} \|_{L^4(\R; L^4)}^2
      \right) 
      \|  U_0(\cdot ) (\varphi_{\lambda}-\varphi_{\lambda'}) \|_{L^4(\R; L^4)}^2 \\
    & \le 
     C\| f \|_{L^{\infty}} \left( 
      \| \varphi_{\lambda}\|_{H^1}^2 + \| \varphi_{\lambda'} \|_{H^1}^2 
      \right) \| \varphi_{\lambda} - \varphi_{\lambda'} \|_{H^1}^2 \\
    & \le 
     C \| f \|_{L^{\infty}} | \lambda - \lambda' |^2.
 \end{align*}
 
 Similarly, for the function $T_G^{(2)}$, one gets  
 \begin{align*}
   T_G^{(2)}(\lambda, \lambda') 
    & \le 
    \| f \|_{L^{\infty}} \int_{\R} 
     \left\| \Cal{F} ( | U_0(t) \varphi_{\lambda} |^2 ) \right\|_{L^2}
     \left\| \Cal{F} ( | U_0(t)\varphi_{\lambda} |^2 - | U_0(t)\varphi_{\lambda'} |^2 ) 
     \right\|_{L^2} \, dt  \\
     & + 
      \| f \|_{L^{\infty}} \int_{\R} 
     \left\| \Cal{F} ( | U_0(t) \varphi_{\lambda'} |^2 ) \right\|_{L^2} 
     \left\| \Cal{F} ( | U_0(t)\varphi_{\lambda} |^2 - | U_0(t)\varphi_{\lambda'} |^2 ) 
     \right\|_{L^2} \, dt  \\
    & \le 
     \| f \|_{L^{\infty}} \int_{\R} 
     \left(  
      \left\| U_0(t) \varphi_{\lambda} \right\|_{L^4}^2 
      +
      \left\| U_0(t) \varphi_{\lambda'} \right\|_{L^4}^2
     \right) \\
     &\hspace{2em} \left\| \, | U_0(t)\varphi_{\lambda} | + | U_0(t)\varphi_{\lambda'} | \right\|_{L^4} \, 
      \left\| \, | U_0(t)\varphi_{\lambda} | -  | U_0(t)\varphi_{\lambda'} | \right\|_{L^4}\, dt\\ 
    & \le 
     \| f \|_{L^{\infty}} \int_{\R} 
      \left(  \| U_0(t) \varphi_{\lambda} \|_{L^4} + \| U_0(t) \varphi_{\lambda'} \|_{L^4}\right)^3 
     \| U_0(t) ( \varphi_{\lambda}-\varphi_{\lambda'} ) \|_{L^4}\, dt \\
    & \le 
     \| f \|_{L^{\infty}} \left( 
      \| U_0(\cdot) \varphi_{\lambda} \|_{L^4(\R; L^4)}^3 
       + \| U_0(\cdot) \varphi_{\lambda'} \|_{L^4(\R; L^4)}^3
      \right) 
      \|  U_0(\cdot ) (\varphi_{\lambda}-\varphi_{\lambda'}) \|_{L^4(\R; L^4)} \\
    & \le 
    C \| f \|_{L^{\infty}} \left( 
      \| \varphi_{\lambda}\|_{H^1}^3 + \| \varphi_{\lambda'} \|_{H^1}^3
      \right) \| \varphi_{\lambda} - \varphi_{\lambda'} \|_{H^1} \\
    & \le 
     C \| f \|_{L^{\infty}} | \lambda - \lambda' |.
 \end{align*}
 Thus we obtain 
 \[
    | (T_G)(f)(\lambda) - (T_G f)(\lambda') | \le C \| f \|_{L^{\infty}}
     | \lambda - \lambda' | 
 \]
 for $ \lambda \in \Gamma$. 
 This implies that $ \{ T_G f \}$ is equicontinuous  and 
 equibounded for $ \| f \|_{L^{\infty}}$.  
 It therefore follows from the theorem Ascoli that $\{ T_G f \}$ contains a Cauchy subsequence 
 in $C(\Gamma)$, which implies 
the integral operator $T_G$ is a  compact operator from 
$L^{\infty}(\R^n)$ to $C(\Gamma)$.   
The proof is complete.

 \hfill $\Box$

\subsection{Integral operator $T_{H}$}\label{sub3-2}
Consider the integral operator $T_H$: 
  \begin{equation}\label{eqn:IEQ3-2}
     (T_H f )(\lambda) 
    :=\int_{\R^n} f(\xi) H(\xi, \lambda)\, d\xi.
  \end{equation}
where 
  \begin{align*}
   H(\xi, \lambda) =  \sum_{\substack{k=1 \\ k \not=j}}^N 
                \int_{\R}  \Cal{F} \left( \left| U_0 (t) \varphi_{\lambda}^{(k)} \right|^2 \right) (\xi) 
                \overline{\Cal{F} \left( \left| U_0 (t) \varphi_{\lambda}^{(j)}  \right|^2 \right) (\xi)} \, dt
  \end{align*}
and 
$\varphi^{(j)}_{\lambda}(x) = \varphi^{(j)}(\, (\lambda +1)x)$.

 \begin{thm}\label{thm:H3-2}
   Let $\Gamma \subset \R$ be a compact set. 
   Assume that $2\le n \le 6$. 
   Then for any $\varphi_j \in H^1(\R^n)$, $j=1,\cdots, N$ the integral operator $T_H$ 
 is a compact operator from $L^{\infty}(\R^n)$ to $C(\Gamma)$. 
 \end{thm}

{\it Proof. }
It is clear that 
  \begin{align*}
    | H(\xi, \lambda) - H(\xi, \lambda') | 
    & \le 
    \sum_{\substack{k=1 \\ k \not=j}}^N  \left| 
                \int_{\R}  
                \Cal{F} \left( \left| U_0 (t) \varphi_{\lambda}^{(k)} \right|^2 \right) (\xi) 
                \overline{\Cal{F} \left( \left| U_0 (t) \varphi_{\lambda}^{(j)}  \right|^2 \right) (\xi)}
                \right.  \\
    & \hspace{2em} - \left.
                \Cal{F} \left( \left| U_0 (t) \varphi_{\lambda'}^{(k)} \right|^2 \right) (\xi) 
                \overline{\Cal{F} \left( \left| U_0 (t) \varphi_{\lambda'}^{(j)}  \right|^2 \right) (\xi)} \, dt 
                \right| \\
    & \le 
     \sum_{\substack{k=1 \\ k \not=j}}^N 
     \left( H^{(1)}(\lambda, \lambda') + H^{(2)}(\lambda, \lambda') \right),
  \end{align*}
where
  \begin{align*}
    H^{(1)}(\xi, \lambda, \lambda') 
     &= \int_{\R} 
     \left| \overline{  \Cal{F} \left( \left| U_0 (t) \varphi_{\lambda}^{(j)} \right|^2 \right) (\xi) }\right|  \\
      &\left| \left\{ 
         \Cal{F} \left( \left| U_0 (t) \varphi_{\lambda}^{(k)} \right|^2 \right) (\xi)  - 
           \Cal{F} \left( \left| U_0 (t) \varphi_{\lambda'}^{(k)} \right|^2 \right) (\xi) 
      \right\} \right| \, dt \\ 
    H^{(2)}(\xi, \lambda, \lambda') 
      &= \int_{\R}   \left| 
      \Cal{F} \left( \left| U_0 (t) \varphi_{\lambda'}^{(k)} \right|^2 \right) (\xi) \right|  \\
      & \left| \left\{ 
        \overline{ \Cal{F} \left( \left| U_0 (t) \varphi_{\lambda}^{(j)} \right|^2 \right) (\xi)}  - 
        \overline{  \Cal{F} \left( \left| U_0 (t) \varphi_{\lambda'}^{(j)} \right|^2 \right) (\xi)} 
      \right\} \right| \, dt .\\ 
  \end{align*}
The same technique as in the estimate on $T_G^{(2)}(\lambda, \lambda')$ finds that 
 \begin{align*}
  & \int_{\R^n} |f(\xi)| \,   | H(\xi, \lambda) - H(\xi, \lambda') | \, d\xi  \\
   & \hspace{6em} \le 
    \| f \|_{L^{\infty}} \sum_{k=1, k\not=j}^N 
    \left\{ c_j \| \varphi_{\lambda}^{(k)} - \varphi_{\lambda'}^{(k)} \|_{H^1} + 
     c_k \| \varphi_{\lambda}^{(j)} - \varphi_{\lambda'}^{(j)} \|_{H^1} 
     \right\}
 \end{align*}
 for some $c_j, c_k>0$. 
 This estimate implies that 
  \[
    | (T_H)(f)(\lambda) - (T_H f)(\lambda') | \le C \| f \|_{L^{\infty}} 
     | \lambda - \lambda' | 
 \]
 for $ \lambda \in \Gamma$. 
 Due to the same argument as in 
 the subsection \ref{sub3-1},  
 the integral operator $T_H$ is the compact operator from 
 $L^{\infty}(\R^n)$ to $C(\Gamma)$. 
 The proof is complete. 
\hfill $\Box$

\subsection{Integral operator $T_{HF}$}\label{sub3-3}
 Consider the integral operator $T_{HF}$: 
  \begin{equation}\label{eqn:IEQ3-3}
    (T_{HF} f )(\lambda) 
    :=\int_{\R^n} f(\xi) H_{HF}(\xi, \lambda)\, d\xi.
  \end{equation}
where
  \begin{align*}
   H_{HF}(\xi, \lambda) &= \sum_{k=1}^N 
                \int_{\R}  \Cal{F} \left(  \left| U_0 (t) \varphi_{\lambda}^{(k)} \right|^2 \right) (\xi) 
                \overline{\Cal{F} \left( \left| U_0 (t) \varphi_{\lambda}^{(j)} \right|^2 \right) (\xi)} \, dt \\
             &\hspace{2em}- \sum_{k=1}^N \int_{\R}  \left| 
                 \Cal{F} \left(  \left(U_0(t) \varphi_{\lambda}^{(j)} \right) 
                \overline{ \left( U_0(t) \varphi_{\lambda}^{(k)} \right) } \right) (\xi) \right|^2
                 \, dt
\end{align*}
 and $\varphi^{(j)}_{\lambda}(x) = \varphi^{(j)}(\, (\lambda +1)x)$.

 \begin{thm}\label{thm:HF3-3}
   Let $\Gamma \subset \R$ be a compact set. 
   Assume that $2\le n \le 6$. 
   Then for any $\varphi_j \in H^1(\R^n)$, $j=1,\cdots, N$ the integral operator $T_{HF}$ 
 is a compact operator from $L^{\infty}(\R^n)$ to $C(\Gamma)$. 
 \end{thm}

 {\it Proof. } 
 We write 
   \begin{align*}
    H_{HF}(\xi, \lambda) - H_{HF}(\xi, \lambda') 
    & = \sum_{k=1}^N 
     \left( H_{HF}^{(1)}(\xi, \lambda, \lambda') - H_{HF}^{(2)}(\xi, \lambda, \lambda') \right),
   \end{align*}
where 
   \begin{align*}
    H_{HF}^{(1)}(\xi, \lambda, \lambda') 
    &=
                 \int_{\R} 
                 \left\{  \Cal{F} \left(  \left| U_0 (t) \varphi_{\lambda}^{(k)} \right|^2 \right) (\xi) 
                \overline{\Cal{F} \left( \left| U_0 (t) \varphi_{\lambda}^{(j)} \right|^2 \right) (\xi)} \right. \\
    & \hspace{2em} - \left. 
                   \Cal{F} \left(  \left| U_0 (t) \varphi_{\lambda'}^{(k)} \right|^2 \right) (\xi) 
                \overline{\Cal{F} \left( \left| U_0 (t) \varphi_{\lambda'}^{(j)} \right|^2 \right) (\xi)}
                \right\} \, dt, \\
    H_{HF}^{(2)}(\xi, \lambda, \lambda') 
    &= 
       \int_{\R}  \left\{
      \left| \Cal{F} \left(  \left(U_0(t) \varphi_{\lambda}^{(j)} \right) 
                \overline{ \left( U_0(t) \varphi_{\lambda}^{(k)} \right) } \right) (\xi)\right|^2 \right. \\
    & \hspace{2em} - \left.
       \left| \Cal{F} \left(  \left(U_0(t) \varphi_{\lambda'}^{(j)} \right) 
                \overline{ \left( U_0(t) \varphi_{\lambda'}^{(k)} \right) } \right) (\xi)\right|^2
                \right\} \, dt. \\ 
  \end{align*}
Due to the fact that $|H^{(1)}_{HF}| \le H^{(1)} + H^{(2)}$, where $H^{(1)}$ and 
$H^{(2)}$ are defined in the proof of Theorem \ref{thm:H3-2},  we have 
   \[
    \int_{\R^n} | H^{(1)}_{HF}(\xi, \lambda, \lambda') | \, d\xi \le 
    C | \lambda - \lambda'|
   \]
 for some $C>0$ and for $\lambda \in \Gamma$.   
 
 For $H^{(2)}_{HF}$, thanks to  the inequality 
 $ 
    |g|^2 - |f|^2 \le | \,  g -  f \,  |^2 +2 | f | \, | \,   g  -  f  \, |,
 $  
we have 
 \begin{align*}
  & | H_{HF}^{(2)}(\xi, \lambda, \lambda') |  \\
   & \hspace{1em} \le 
   \int_{\R} \left| 
    \Cal{F} \left(  \left(U_0(t) \varphi_{\lambda}^{(j)} \right)
                \overline{ \left( U_0(t) \varphi_{\lambda}^{(k)} \right) } \right)(\xi) -
    \Cal{F} \left(  \left(U_0(t) \varphi_{\lambda'}^{(j)} \right) 
                \overline{ \left( U_0(t) \varphi_{\lambda'}^{(k)} \right) } \right)(\xi)
    \right|^2 \, dt \\
    &\hspace{2em} + 2
    \int_{\R} 
    \left| \Cal{F} \left(  \left(U_0(t) \varphi_{\lambda'}^{(j)} \right) 
                \overline{ \left( U_0(t) \varphi_{\lambda'}^{(k)} \right) } \right)(\xi) \right| \times \\
    & \hspace{3em} \times 
    \left| \Cal{F} \left(  \left(U_0(t) \varphi_{\lambda}^{(j)} \right) 
                \overline{ \left( U_0(t) \varphi_{\lambda}^{(k)} \right) } \right) (\xi)- 
                \Cal{F} \left(  \left(U_0(t) \varphi_{\lambda'}^{(j)} \right) 
                \overline{ \left( U_0(t) \varphi_{\lambda'}^{(k)} \right) } \right) (\xi)\right| \, dt.
 \end{align*}
  Thanks to Lemma \ref{lem:RH2-4}, one gets  
  \begin{align*}
   &    \int_{\R^n} \int_{\R} \left| 
    \Cal{F} \left(  \left(U_0(t) \varphi_{\lambda}^{(j)} \right) 
                \overline{ \left( U_0(t) \varphi_{\lambda}^{(k)} \right) } \right) -
    \Cal{F} \left(  \left(U_0(t) \varphi_{\lambda'}^{(j)} \right) 
                \overline{ \left( U_0(t) \varphi_{\lambda'}^{(k)} \right) } \right)
    \right|^2 \, dt\, d\xi \\
    & \le 
    \int_{\R} 
     \left\|  \left(U_0(t) \varphi_{\lambda}^{(j)} \right) 
                \overline{ \left( U_0(t) \varphi_{\lambda}^{(k)} \right) } -
       \left(U_0(t) \varphi_{\lambda'}^{(j)} \right) 
                \overline{ \left( U_0(t) \varphi_{\lambda'}^{(k)} \right) } \right\|_{L^2}^2 \, dt \\
     & \le
      \int_{\R} 
      \left\|  \overline{\left(U_0(t) \varphi_{\lambda}^{(k)} \right)} 
                \left\{ \left( U_0(t) \varphi_{\lambda}^{(j)} \right)  -
       \left(U_0(t) \phi_{\lambda'}^{(j)} \right) 
       \right\}
                \right\|_{L^2}^2 \, dt \\ 
      & + 
       \int_{\R} 
      \left\|  \left(U_0(t) \varphi_{\lambda'}^{(j)} \right) 
                \left\{ \overline{ \left( U_0(t) \varphi_{\lambda}^{(k)} \right)}  -
       \overline{\left(U_0(t) \varphi_{\lambda'}^{(k)} \right)} 
       \right\}
                \right\|_{L^2}^2 \, dt \\
      & \le 
       C_1 \| \varphi_{\lambda}^{(k)} \|_{H^1}^2 
        \| \phi_{\lambda}^{(j)} - \varphi_{\lambda'}^{(j)} \|_{H^1}^2 + 
        C_2 \| \varphi_{\lambda'}^{(j)} \|_{H^1}^2 
         \| \varphi_{\lambda}^{(k)} - \varphi_{\lambda'} ^{(k)} \|_{H^1}^2 \\
      &\le C |\lambda - \lambda'| 
  \end{align*}
for some $C>0$ and $\lambda \in \Gamma$.  
In the same way as in the proof of Theorem \ref{thm:3-1},  
we also obtain  
 \begin{align*}
   &\int_{\R^n} \int_{\R} 
    \left| \Cal{F} \left(  \left(U_0(t) \varphi_{\lambda'}^{(j)} \right) 
                \overline{ \left( U_0(t) \varphi_{\lambda'}^{(k)} \right) } \right) \right| \times \\
    & \hspace{2em} \times 
    \left| \Cal{F} \left(  \left(U_0(t) \varphi_{\lambda}^{(j)} \right) 
                \overline{ \left( U_0(t) \varphi_{\lambda}^{(k)} \right) } \right) - 
                \Cal{F} \left(  \left(U_0(t) \varphi_{\lambda'}^{(j)} \right) 
                \overline{ \left( U_0(t) \varphi_{\lambda'}^{(k)} \right) } \right) \right| \, dt \, d\xi \\
    & \le 
        \int_{\R}  \left\|  \left(U_0(t) \varphi_{\lambda'}^{(j)} \right)  
         \overline{   \left(U_0(t) \varphi_{\lambda'}^{(k)} \right) } \right\|_{L^2} \times \\
    & \hspace{2em} 
        \left\|   \left( U_0(t) \varphi_{\lambda}^{(j)} \right)  
           \overline{   \left(U_0(t) \varphi_{\lambda}^{(k)} \right)  } - 
             \left(U_0(t) \varphi_{\lambda'}^{(j)} \right)  
           \overline{   \left(U_0(t) \varphi_{\lambda'}^{(k)} \right)} \right\|_{L^2}\, dt \\
    & \le 
        \int_{\R} 
         \left\{ \left\| \, \left| U_0(t) \varphi_{\lambda'}^{(j)} \right|^2 \right\|_{L^2} + 
         \left\| \, \left| U_0(t) \varphi_{\lambda'}^{(k)} \right|^2 \right\|_{L^2} \right\} \times \\
    & \hspace{2em}
         \left\{ 
            \left\|   \left( U_0(t) \varphi_{\lambda}^{(j)} \right)  \left\{
           \overline{   \left(U_0(t) \varphi_{\lambda}^{(k)} \right)  } -   
           \overline{   \left(U_0(t) \varphi_{\lambda'}^{(k)} \right)} \right\} \right\|_{L^2} \right. \\
    & \hspace{3em}+
            \left. 
            \left\|   \overline{ \left( U_0(t) \varphi_{\lambda'}^{(k)} \right) }  \left\{
             \left(U_0(t) \varphi_{\lambda}^{(j)} \right)   -   
             \left(U_0(t) \varphi_{\lambda'}^{(j)} \right) \right\} \right\|_{L^2}
             \right\}\, dt \\
     & \le 
         \int_{\R} 
           \left\{ \left\|  U_0(t) \varphi_{\lambda'}^{(j)}  \right\|_{L^4}^2 + 
         \left\|  U_0(t) \varphi_{\lambda'}^{(k)}  \right\|_{L^4}^2 \right\} \times \\
    & \hspace{2em}
         \left\{ 
            \left\|    U_0(t) \varphi_{\lambda}^{(j)} \right\|_{L^4}  \left\|
           \overline{   U_0(t) \varphi_{\lambda}^{(k)}  } -   
           \overline{   U_0(t) \varphi_{\lambda'}^{(k)} } \right\|_{L^4} \right. \\
    & \hspace{3em}+
            \left. 
            \left\|   \overline{  U_0(t) \varphi_{\lambda'}^{(k)}  }  \right\|_{L^4} \left\|
             U_0(t) \varphi_{\lambda}^{(j)}    -   
             U_0(t) \varphi_{\lambda'}^{(j)}   \right\|_{L^4}
             \right\}\, dt \\
     & \le 
         \int_{\R} 
           \left( \left\|  U_0(t) \varphi_{\lambda'}^{(j)}  \right\|_{L^4} + 
         \left\|  U_0(t) \varphi_{\lambda'}^{(k)}  \right\|_{L^4}  \right)^3 \times \\
    & \hspace{2em}
         \left( 
            \left\|   
           \overline{   U_0(t) \varphi_{\lambda}^{(k)}  } -   
           \overline{   U_0(t) \varphi_{\lambda'}^{(k)} } \right\|_{L^4} +
             \left\|
             U_0(t) \varphi_{\lambda}^{(j)}    -   
             U_0(t) \varphi_{\lambda'}^{(j)}   \right\|_{L^4}
             \right)\, dt \\
     & \le 
          \left( \left\|  U_0(t) \varphi_{\lambda'}^{(j)}  \right\|_{L^4(\R; L^4)}^3 + 
         \left\|  U_0(t) \varphi_{\lambda'}^{(k)}  \right\|_{L^4(\R; L^4)}^3  \right) \times \\
    & \hspace{2em}
         \left( 
            \left\|   
           \overline{   U_0(t) \varphi_{\lambda}^{(k)}  } -   
           \overline{   U_0(t) \varphi_{\lambda'}^{(k)} } \right\|_{L^4(\R; L^4)} +
             \left\|
             U_0(t) \varphi_{\lambda}^{(j)}    -   
             U_0(t) \varphi_{\lambda'}^{(j)}   \right\|_{L^4(\R; L^4)}
             \right) \\
     & \le C 
         \left( 
           \| \varphi_{\lambda'}^{(j)} \|_{H^1}^3 
           + \| \varphi_{\lambda'}^{(k)} \|_{H^1}^3 
        \right) 
        \left( 
         \| \varphi_{\lambda}^{(k)} - \varphi_{\lambda'}^{(k)} \|_{H^1}
         + 
        \| \varphi_{\lambda}^{(j)} - \varphi_{\lambda'}^{(j)} \|_{H^1} 
       \right) \\
    & \le C | \lambda - \lambda' |
  \end{align*}
for some $C>0$ and $\lambda \in \Gamma$. 

Consequently,  we obtain
\[
 \int_{\R^n} \left| H^{(2)}(\xi, \lambda, \lambda') \right|\, d\xi 
  \le C | \lambda - \lambda'|
\]
for some $C>0$ and $\lambda \in \Gamma$. Hence one gets  
\begin{align*}
 \left| (T_{HF} f)(\lambda) - (T_{HF}f)(\lambda') \right| 
  & \le 
   \| f \|_{L^{\infty}} \int_{\R^n} 
    \left| H_{HF}(\xi, \lambda) - H_{HF}(\xi, \lambda') \right| \, d\xi \\
  & \le \| f \|_{L^{\infty}} \sum_{k=1}^N 
   \int_{\R^n} \left| H_{HF}^{(1)}(\xi, \lambda, \lambda') \right| + 
    \left| H_{HF}^{(2)}(\xi, \lambda, \lambda') \right| \, d\xi \\
  & \le C \| f \|_{L^{\infty}} | \lambda - \lambda'|   
\end{align*}
for some $C>0$ and $\lambda \in \Gamma$.  
This implies that  
the operator $T_{HF}$ is a compact operator from $L^{\infty}(\R^n)$ to 
$C(\Gamma)$, due to 
the same argument as in the proof of Theorem \ref{thm:3-1}. 
The proof is complete. 
 \hfill $\Box$


\section{Uniqueness}\label{sec:4}
In this section, we prove the following uniqueness theorem in the 
inverse scattering problem for the HF equation \eqref{eqn:1-2}. 

\begin{thm}\label{thm:4-1}
Let $2\le n \le 6$. 
Assume that $V_{\sharp}, \sharp = 1,2$ satisfy the assumption in 
Theorem \ref{thm:HF-2}. 
Let $S_{\sharp}$ are the scattering operator for the HF equation 
\eqref{eqn:1-2} with interaction potential $V_{\sharp}$. 
If $S_1=S_2$, then we have $V_1=V_2$. 
\end{thm}

\subsection{Lemmas}
In order to prove theorem \ref{thm:4-1}, 
we need some lemmas. 
Set 
\begin{align*}
  ( \Cal{G}_1 \varphi^{(k)} ) (t,\xi) 
  &= \Cal{F}\left( \left| e^{-itH_0}\varphi^{(k)} 
  \right| ^2 \right) (\xi), \\
  ( \Cal{G}_2 [\varphi^{(j)}, \varphi^{(k)}] ) (t,\xi) 
  &= \Cal{F}\left(  e^{-itH_0}\varphi^{(j)} \overline{e^{-itH_0}\varphi^{(k)}} 
   \right) (\xi) \\
\end{align*}
and put $B_R(a) =\{ x\in \R^n\, ; \, |x-a| < R\}$. 

\begin{lem}\label{lem:4-1}
 Let $\ve >0$ and $p_k \in \R^n$ such that $\cap_{k=1}^N B_{\ve}(p_k) =\emptyset$. 
 Then for any $\varphi^{(k)} \in \Cal{S}_0$ with 
 $\displaystyle{\dai_{\xi} \widehat{\varphi^{(k)}}(\xi) \subset B_{\ve}(p_k), 
 k=1,2, \cdots N}$, we have 
  \[
    ( \Cal{G}_1\varphi^{(k)}) (t,\xi) \overline{ ( \Cal{G}_1\varphi^{(j)}) (t,\xi)}=0, \quad 
     k\not= j, 
  \]
 on $(t,\xi) \in \R \times \R^n$. 
\end{lem}

{\it Proof. }
 Due to the fact that $\widehat{|f|^2}=\widehat{f\overline{f}}= \widehat{f}*\widehat{\overline{f}}$, 
 we have 
 \begin{align}\label{eqn:4-1}
   \notag ( \Cal{G}_1 \varphi^{(k)} )(t, \xi) 
   &=
    e^{-it\xi^2} \widehat{\varphi^{(k)}} * e^{it\xi^2} \widehat{\overline{\varphi^{(k)}}} \\ \notag
  &= 
    \int_{\R^n} e^{-it(\xi-\eta)^2} \widehat{\varphi^{(k)}}(\xi - \eta) 
     e^{it\eta^2} \widehat{\overline{\varphi^{(k)}}}(\eta) \, d\eta \\ 
  &=
    \int_{\R^n} e^{-it \xi^2} e^{2it\xi \cdot \eta}  \widehat{\varphi^{(k)}}(\xi - \eta) 
    \widehat{\overline{\varphi^{(k)}}}(\eta) \, d\eta. 
 \end{align}
 Note that if $\dai_{\eta} \widehat{\varphi^{(k)}}(\eta) \subset B_{\ve}(p_k)$, 
 then $\dai_{\xi} \widehat{\varphi^{(k)}}(\xi - \eta) \widehat{\varphi^{(k)}}(\eta) 
 \subset B_{2\ve} (2p_k)$ for each $\eta \in B_{\ve}(p_k)$.  
 It is also clear that if $ \cap_{k=1}^N B_{\ve}(p_k) = \emptyset$, 
 then $\cap_{k=1}^N B_{2\ve}(2p_k) = \emptyset$. 
 Thus, one has 
  \[
     \cap_{k=1}^N \dai_{\xi} \left( \widehat{\varphi^{(k)}}(\xi - \eta) \widehat{\varphi^{(k)}}(\eta) 
     \right) = \emptyset
  \] 
 for $\eta \in B_{\ve}(p_k)$, which implies that 
 $\dai_{(t,\xi)} \Cal{G}_1 \varphi^{(k)} \cap \dai_{(t,\xi)} \Cal{G}_1 \varphi^{(j)} = \emptyset$, 
 due to the identity \eqref{eqn:4-1}, we complete the proof. 
\hfill $\Box$

\begin{lem}\label{lem:4-2}
 Let $\ve >0$ and $p_{\ell} \in \R^n, \ell=1,2, \cdots, N$ such that 
 $\cap_{\ell=1}^N B_{\ve}(p_{\ell})= \emptyset$. 
 Then for any $ \varphi^{(\ell)} \in \Cal{S}_0$ with 
 $\dai_{\xi} \widehat{\varphi^{(\ell)}}(\xi) \subset B_{\ve}(p_{\ell})$, we have 
 \[
    \dai_{(t,\xi)} (\Cal{G}_2 [ \varphi^{(j)}, \varphi^{(k)} ])(t, \xi) \subset 
    \R \times B_{2\ve}(p_j + p_k), \qquad 1\le j \not=k \le N.
 \]
\end{lem}

{\it Proof. }
 It is clear that 
  \[
      ( \Cal{G}_2[ \varphi^{(j)}, \varphi^{(k)}] )(t, \xi) 
   =
       e^{-it \xi^2}  \int_{\R^n}e^{2it\xi \cdot \eta}  \widehat{\varphi^{(j)}}(\xi - \eta) 
    \widehat{\overline{\varphi^{(k)}}}(\eta) \, d\eta. 
  \]
In view of the assumption  $\dai_{\xi} \widehat{\varphi^{(\flat)}}(\xi) \subset B_{\ve}(p_{\flat})$, 
$\flat=j,k$, we have 
 \[
     \dai_{\xi} \widehat{\varphi^{(j)}}(\xi-\eta)
      \widehat{\overline{\varphi^{(k)}}}(\eta) \subset B_{2\ve}(p_j + p_k), 
 \]
which proves Lemma \ref{lem:4-2}.  
\hfill $\Box$

\subsection{Proof of Theorem \ref{thm:4-1}}
We are now in a position to prove Theorem \ref{thm:4-1}.

Let $w(\xi)= \widehat{V}_1(\xi) - \widehat{V}_2(\xi)$. 
By virtue of Theorem \ref{thm:HF-2} and assumption $S_1=S_2$, one has 
 \begin{align*}
   0&=
    \int_{\R^n} w(\xi) H^{(j)}_{HF} (\xi) \, d\xi \\
    &= \int_{\R^n} w(\xi) \sum_{k=1}^N \int_{\R} 
      (\Cal{G}_1 \varphi^{(k)})(t,\xi) \overline{(\Cal{G}_1 \varphi^{(j)})(t, \xi)}\, dt  \\
    &\hspace{1em} - 
     \int_{\R^n} w(\xi) \sum_{k=1}^N \int_{\R} 
      \left| (\Cal{G}_2 [\varphi^{(k)}, \varphi^{(j)}])(t, \xi) \right|^2\, dt
 \end{align*}
for any $\varphi^{(\ell)}\in \Cal{S}_0, \ell=1,2, \cdots, N$. 
Taking   $\varphi^{(\ell)}\in \Cal{S}_0$ with 
$\dai_{\xi} \widehat{\varphi^{(\ell)}}(\xi) \subset B_{\ve}(p_{\ell})$ 
as in Lemma \ref{lem:4-1}, we have 
\[
   \int_{\R^n} w(\xi) \sum_{k=1}^N \int_{\R} 
      (\Cal{G}_1 \varphi^{(k)})(t,\xi) \overline{(\Cal{G}_1 \varphi^{(j)})(t, \xi)}\, dt =0.
\] 
This implies that for such functions $\varphi^{(\flat)}, \flat=j,k$
 \begin{align}\label{eqn:4-2}
    & \int_{\R^n} \mathrm{Re} \left( w(\xi) \right) \sum_{k=1}^N \int_{\R} 
      \left| (\Cal{G}_2 [\varphi^{(k)}, \varphi^{(j)}])(t, \xi) \right|^2\, dt =0, \\ \notag
    & \int_{\R^n} \mathrm{Im} \left( w(\xi) \right) \sum_{k=1}^N \int_{\R} 
      \left| (\Cal{G}_2 [\varphi^{(k)}, \varphi^{(j)}])(t, \xi) \right|^2\, dt =0, 
 \end{align}
 where  $\mathrm{Re} \left( w(\xi) \right)$ and $\mathrm{Im} \left( w(\xi) \right)$  denote the real part and the imaginary part of the complex-valued function $w(\xi)$, respectively.  
 
 Assume that for $m>n/2$, $ H^m(\R^n) \ni w \not\equiv 0$ on $\R^n$. 
 Due to the Sobolev embedding theorem, 
 we find  $w\in C^{m-[n/2]-1, \gamma}(\R^n)$ for some $0<\gamma <1$.  
 Here $C^{m, \gamma}(\R^n)$ denotes the set of H\"{o}lder continuous functions 
 with exponent $\gamma$. 
 Therefore, there exist $\delta>0$ and $p\in \R^n$ such that 
 $\mathrm{Re}\left( w(\xi) \right) >0$ on $B_{\delta}(p)$.  
 For $\ve >0$ sufficiently small and fixed $p_j\in \R^n$, we define $p_{\ell}, \ell=1,2, \cdots, N$ so that 
 $p_{\ell} \not= p_j, 1\le \ell, j \le N$ and 
 \[
    \cup_{\ell=1, \ell \not=j}^N B_{2\ve}(p_{\ell}+p_j) \subset B_{\delta}(p).
 \]
 Then, thanks to  Lemma \ref{lem:4-2}, there exist  $\varphi^{(\flat)}\in \Cal{S}_0$, $\flat=j,k$ 
 such that the function 
 $\sum_{k=1}^N(\int | \Cal{G}_2[ \varphi^{(k)}, \varphi^{(j)}]|^2 dt)(\xi)$ is a non-negative function 
 with  the support in $B_{\delta}(p)$. 
 This and the positivity $\mathrm{Re}\left( w(\xi) \right)>0$ on 
 $B_{\delta}(p)$ show that 
 the integral of the left hand side in the identity \eqref{eqn:4-2} never vanishes 
 for such functions $\varphi^{(\flat)}$. 
 This contradicts the identity  \eqref{eqn:4-2}. 
 Thus, we conclude that $w \equiv 0$ on $\R^n$, which proves Theorem \ref{thm:4-1}.



%
%

%

\bibliographystyle{plain}


\begin{thebibliography}{10}

\bibitem{Cazenave2003}
Thierry Cazenave.
\newblock {\em Semilinear {S}chr\"{o}dinger equations}, volume~10 of {\em
  Courant Lecture Notes in Mathematics}.
\newblock New York University, Courant Institute of Mathematical Sciences, New
  York; American Mathematical Society, Providence, RI, 2003.

\bibitem{Enss-Weder1995}
Volker Enss and Ricardo Weder.
\newblock The geometrical approach to multidimensional inverse scattering.
\newblock {\em J. Math. Phys.}, 36(8):3902--3921, 1995.

\bibitem{goeke-reinhard}
K.~Goeke and P.~G. Reinhard, editors.
\newblock {\em Time-dependent Hartree-Fock and beyond}, volume 171 of {\em
  Lecture notes in physics}.
\newblock Springer-Verlag, Jan 1982.

\bibitem{isozaki1983}
Hiroshi Isozaki.
\newblock On the existence of solutions of time-dependent {H}artree-{F}ock
  equations.
\newblock {\em Publ. Res. Inst. Math. Sci.}, 19(1):107--115, 1983.

\bibitem{isozaki-nakazawa-uhlmann}
Hiroshi Isozaki, Hideo Nakazawa, and Gunther Uhlmann.
\newblock Inverse scattering problem in nuclear physics---optical model.
\newblock {\em J. Math. Phys.}, 45(7):2613--2632, 2004.

\bibitem{kramer-saraceno}
Peter Kramer and Marcos Saraceno.
\newblock {\em Geometry of the time-dependent variational principle in quantum
  mechanics}, volume 140 of {\em Lecture Notes in Physics}.
\newblock Springer-Verlag, Berlin-New York, 1981.

\bibitem{kress}
Rainer Kress.
\newblock {\em Linear integral equations}, volume~82 of {\em Applied
  Mathematical Sciences}.
\newblock Springer-Verlag, New York, second edition, 1999.

\bibitem{mochizuki}
Kiyoshi Mochizuki.
\newblock On small data scattering with cubic convolution nonlinearity.
\newblock {\em J. Math. Soc. Japan}, 41(1):143--160, 1989.

\bibitem{nakanishi1999}
Kenji Nakanishi.
\newblock Energy scattering for {H}artree equations.
\newblock {\em Math. Res. Lett.}, 6(1):107--118, 1999.

\bibitem{novikov}
Roman~G. Novikov.
\newblock On inverse scattering for the {$N$}-body {S}chr\"{o}dinger equation.
\newblock {\em J. Funct. Anal.}, 159(2):492--536, 1998.

\bibitem{Angeles-Romero-Weder2006}
Sandoval Romero, Mar\'{i}a de~los \'{A}ngeles, and Ricardo Weder.
\newblock The initial value problem, scattering and inverse scattering, for
  {S}chr\"{o}dinger equations with a potential and a non-local nonlinearity.
\newblock {\em J. Phys. A}, 39(37):11461--11478, 2006.

\bibitem{sasaki-watanabe}
Hironobu. Sasaki and Michiyuki. Watanabe.
\newblock Uniqueness on identification of cubic convolution nonlinearity.
\newblock {\em J. Math. Anal. Appl.}, 309(1):294--306, 2005.

\bibitem{Strauss1974}
Walter Strauss.
\newblock Nonlinear scattering theory.
\newblock In LaVita~J. A. and Marchand J.-P., editors, {\em Scattering Theory
  in Mathematical Physics}, volume~9 of {\em Nato Advanced Study Institutes
  Series C}, pages 53--78. D. Reidel, 1974.

\bibitem{uhlmann-vasy}
Gunther Uhlmann and Andr\'{a}s Vasy.
\newblock Low-energy inverse problems in three-body scattering.
\newblock {\em Inverse Problems}, 18(3):719--736, 2002.

\bibitem{vasy}
Andr\'{a}s Vasy.
\newblock Structure of the resolvent for three-body potentials.
\newblock {\em Duke Math. J.}, 90(2):379--434, 1997.

\bibitem{wada}
Takeshi Wada.
\newblock Scattering theory for time-dependent {H}artree-{F}ock type equation.
\newblock {\em Osaka J. Math.}, 36(4):905--918, 1999.

\bibitem{wang}
X.~P. Wang.
\newblock On the uniqueness of inverse scattering for {$N$}-body systems.
\newblock {\em Inverse Problems}, 10(3):765--784, 1994.

\bibitem{watanabe0}
Michiyuki Watanabe.
\newblock Inverse scattering for the nonlinear {S}chr\"{o}dinger equation with
  cubic convolution nonlinearity.
\newblock {\em Tokyo J. Math.}, 24(1):59--67, 2001.

\bibitem{watanabe2}
Michiyuki Watanabe.
\newblock Uniqueness in the inverse scattering problem for {H}artree type
  equation.
\newblock {\em Proc. Japan Acad. Ser. A Math. Sci.}, 77(9):143--146, 2001.

\bibitem{watanabe2002}
Michiyuki Watanabe.
\newblock Reconstruction of the {H}artree-type nonlinearity.
\newblock {\em Inverse Problems}, 18(6):1477--1481, 2002.

\bibitem{watanabe2007}
Michiyuki Watanabe.
\newblock Inverse scattering problem for time dependent {H}artree-{F}ock
  equations in the three-body case.
\newblock {\em J. Math. Phys.}, 48(5):053510, 9, 2007.

\bibitem{watanabe2007-1}
Michiyuki Watanabe.
\newblock A remark on inverse scattering for time dependent hartree equations.
\newblock In {\em Journal of Physics: Conference Series}, volume~73, page
  012025, 2007.

\bibitem{watanabe18}
Michiyuki Watanabe.
\newblock Time-dependent method for non-linear {S}chr\"{o}dinger equations in
  inverse scattering problems.
\newblock {\em J. Math. Anal. Appl.}, 459(2):932--944, 2018.

\bibitem{Weder1997}
Ricardo Weder.
\newblock Inverse scattering for the nonlinear {S}chr\"{o}dinger equation.
\newblock {\em Comm. Partial Differential Equations}, 22(11-12):2089--2103,
  1997.

\bibitem{Weder2000_2}
Ricardo Weder.
\newblock Inverse scattering on the line for the nonlinear {K}lein-{G}ordon
  equation with a potential.
\newblock {\em J. Math. Anal. Appl.}, 252(1):102--123, 2000.

\bibitem{Weder2000_1}
Ricardo Weder.
\newblock {$L^p$}-{$L^{\dot p}$} estimates for the {S}chr\"{o}dinger equation
  on the line and inverse scattering for the nonlinear {S}chr\"{o}dinger
  equation with a potential.
\newblock {\em J. Funct. Anal.}, 170(1):37--68, 2000.

\bibitem{Weder2001_2}
Ricardo Weder.
\newblock Inverse scattering for the non-linear {S}chr\"{o}dinger equation:
  reconstruction of the potential and the non-linearity.
\newblock {\em Math. Methods Appl. Sci.}, 24(4):245--254, 2001.

\bibitem{Weder2001_1}
Ricardo Weder.
\newblock Inverse scattering for the nonlinear {S}chr\"{o}dinger equation.
  {II}. {R}econstruction of the potential and the nonlinearity in the
  multidimensional case.
\newblock {\em Proc. Amer. Math. Soc.}, 129(12):3637--3645, 2001.

\bibitem{Weder2002}
Ricardo Weder.
\newblock Multidimensional inverse scattering for the nonlinear
  {K}lein-{G}ordon equation with a potential.
\newblock {\em J. Differential Equations}, 184(1):62--77, 2002.

\bibitem{Weder2005_1}
Ricardo Weder.
\newblock Scattering for the forced non-linear {S}chr\"{o}dinger equation with
  a potential on the half-line.
\newblock {\em Math. Methods Appl. Sci.}, 28(10):1219--1236, 2005.

\end{thebibliography}


\end{document}